\documentclass[final,5p,times,twocolumn]{elsarticle}
\usepackage{amssymb}
\usepackage{textcomp}
\usepackage{amsmath}
\usepackage{booktabs}
\usepackage{geometry}
\usepackage[figuresright]{rotating}
\usepackage{cmap}
\usepackage{hyperref}
\usepackage{xcolor}
\definecolor{color}{HTML}{0000FF}
\hypersetup{linkcolor=color,urlcolor=color,citecolor=color,filecolor=color,colorlinks=true,pdfauthor=author}
\journal{Advances in Colloid and Interface Science}
\begin{document}
\begin{frontmatter}

\title{Applying Droplets and Films in Evaporative Lithography}

\author[label1,label2,label3]{K.S.~Kolegov}
\author[label3]{L.Yu.~Barash\corref{cor}}
\ead{barash@itp.ac.ru}
\address[label1]{Astrakhan State University, 414056 Astrakhan, Russia}
\address[label2]{Volga State University of Water Transport,
Caspian Institute of Maritime and River Transport, 414014 Astrakhan, Russia}
\address[label3]{Landau Institute for Theoretical Physics, 142432 Chernogolovka, Russia}
\cortext[cor]{Corresponding author.}

\begin{abstract}
This review covers experimental results of evaporative lithography
and analyzes existing mathematical models of this method.
Evaporating droplets and films are used in different fields, such
as cooling of heated surfaces of electronic devices, diagnostics in
health care, creation of transparent conductive coatings on flexible
substrates, and surface patterning. A method called evaporative
lithography emerged after the connection between the coffee ring effect
taking place in drying colloidal droplets and naturally occurring
inhomogeneous vapor flux densities from liquid--vapor interfaces was
established. Essential control of the colloidal particle deposit
patterns is achieved in this method by producing ambient conditions
that induce a nonuniform evaporation profile from the colloidal
liquid surface. Evaporative lithography is part of a wider
field known as ``evaporative-induced self-assembly'' (EISA). EISA
involves methods based on contact line processes, methods employing
particle interaction effects, and evaporative lithography. As a rule,
evaporative lithography is a flexible and single-stage process with such
advantages as simplicity, low price, and the possibility of application to
almost any substrate without pretreatment. Since there is no
mechanical impact on the template in evaporative lithography, the
template integrity is preserved in the process. The method is also
useful for creating materials with localized functions, such as
slipperiness and self-healing. For these reasons, evaporative lithography
attracts increasing attention and has a number of noticeable achievements
at present. We also analyze limitations of the approach and ways
of its further development.
\end{abstract}


\end{frontmatter}

\section{Introduction}
\label{sec:intro}

The drying of droplets and films on solid surfaces is accompanied by
a number of phenomena of general physical interest. As a result of
evaporation, the deposition patterns of particles contained in the
liquid can be spatially structured in one or another way depending on
the experimental conditions. This process has proved to be
technologically involved in many applied studies and has led to the
development of the evaporative lithography method addressed in this
review.

Before proceeding to the main subject of the review, we briefly
describe the basic physical processes that occur in evaporating
droplets and films. These open systems exchange mass and energy with
the environment, and their complex behavior is due to many factors.
Various physical and chemical properties of the liquid solution,
substrate, and ambient air as well as external forces affect the
evaporation mode.

Of the many physical processes occurring in droplet and film systems,
the basic process is droplet evaporation into the surrounding gas,
which since Maxwell's studies has long been regarded as mainly vapor
diffusion from the droplet surface (also see~\cite{Maxwell1877,Langmuir1918,Fuchs1959}).
The classic quasistationary theory of droplet evaporation does not
include the effects of fluid dynamics in the droplet and only
partially takes the simplest elements of heat transfer into account.

In the last twenty years, the evaporation of droplets and films has
attracted renewed attention due to the development of new
applications such as the evaporation and combustion of fuel droplets in
engines~\cite{Sazhin2018_6498,Gambaryan-Roisman2019_115}, the
interaction of droplets with surfaces of varying wettability in jet
printing~\cite{Shikhmurzaev2012_082001}, the deposition of microarrays
of DNA and RNA molecules~\cite{Dugas2005,Maheshwari2008}, pattern
formation on surfaces~\cite{Larson20141538,Sefiane2014}, disease
diagnostics techniques~\cite{Trantum2011}, the creation of photonic
crystals~\cite{Masuda20054478,Wang20136048,Zhao2014}, the production of
nanoparticles~\cite{Fisenko2014599}, the production of stable ultrathin
film coatings of polar liquids~\cite{Gordeeva2017_531}, cooling of
devices~\cite{Kabov2007103,Gatapova20084797,Kabov2011825}, removal
of nanoparticles from a solid surface for cleaning
purposes~\cite{MAHDI201513}, etc.

The problem of the evaporation of droplets and films containing
nanoparticles aroused a large interest. Nanoparticle structures can
appear on the free surface during evaporation and also remain on the
substrate after drying. One of the examples is the self-assembly of
nanoparticles into organized superlattices (see, e.g.,~\cite{Bigioni2006}).

The presence of substrate nonuniformity and defects normally causes
effective fixation (pinning) of the three-phase boundaries of sessile
droplets. Observation of different evaporation process
stages~\cite{Picknett1977,Shanahan1994,Bourges1995,Kajiya2006} has
shown that the longest and dominant regime of the evaporation
process is where the contact line is pinned and the contact
area of a drop with a solid base remains constant. As the volume of
the sessile droplet decreases, periodic displacements and fixations
of the contact line can occur. As the contact line further becomes
completely depinned, a different regime, the constant contact
angle mode, switches on. In the final drying stage, the height, the
contact area, and the contact angle rapidly decrease with time.

Among the processes of spatial structuring of deposition patterns
from drying sessile droplets, the best known is the so-called
coffee-ring effect, which is observed in evaporating droplets of
colloidal and molecular solutions if the deposits are formed near
the pinned contact line. The studies of Deegan et al. theoretically
proved that mass loss resulting from vaporization occurs
nonuniformly along the free surface of the liquid layer,
significantly increasing near the contact
line~\cite{Deegan1997,Deegan2000,Deegan2000475}. Because the
evaporation rate dominates at the droplet periphery, compensatory
flows are generated that transfer suspended and dissolved matter to
the three-phase boundary, resulting in an annular distribution of
deposited particles near the contact line. The number of concentric
structures of particles formed on the substrate depends on the
number of periodic modes of displacement and fixation of the
contact line, which can occur with a decrease in the droplet
volume~\cite{Deegan2000475}. The behavior of the three-phase
boundary and the type of annular deposition, including the presence of
cracks, depend on the substrate roughness and particle
size~\cite{Lohani2020}.

A nonuniform mass flow during evaporation and the corresponding heat
transfer result in more than just a compensatory flow. They also change the
temperature distribution along the droplet surface, and because the
surface tension depends on temperature, this can cause Marangoni forces.
These forces produce thermocapillary convection inside a droplet~\cite{HuLarson2005,Barash2009},
which differs qualitatively from classical Marangoni convection studied in
systems with a simple flat geometry~\cite{Benard1900, Pearson1958}. Under
certain conditions, the value of the thermal conductivity of the substrate
determines the tangential component of the temperature gradient at the
droplet surface near the contact line and consequently affects the direction
of convection inside the droplet~\cite{Ristenpart2007,Barash2015}.

In the presence of thermocapillary flows, some of the colloidal particles are
sometimes deposited in the central part of the droplet, thus forming a
spot inside the ring~\cite{Baek2018477}. A uniform distribution of the
deposit over the entire area where the liquid had previously come into
contact with the substrate can be due to the effect of the capture of
colloidal particles by the descending free surface~\cite{Li201624628}.
A similar pattern is obtained when a surfactant and a surface-absorbed
polymer are added to a binary mixture~\cite{Kim2016124501}. When salt is
added, its crystallization is observed in the deposit in some
cases~\cite{Ragoonanan2008,Yakhno2008225,Morinaga2020}. More details about
methods for suppressing the coffee-ring effect to obtain uniform deposits
can be found in~\cite{Mampallil201838}.

For rather large contact angles $\theta$, the thermocapillary flow can
significantly exceed the compensatory flow. But with small $\theta$, the
compensatory flow dominates the Marangoni convection for the following
reasons. First, the Marangoni flow velocity is proportional to $\theta^2$
and decreases rapidly in the final stage of the evaporation process.
Second, the thermocapillary instability is known to occur only when the Marangoni
number $\mathrm{Ma}=-\sigma^\prime\Delta Th_0/(\eta\kappa)$ exceeds a
certain critical value. Here, $\Delta T$ is the temperature difference
between the apex and the substrate, $\sigma^\prime$ is the
temperature-based derivative of the surface tension, $h_0$ is the droplet
height, $\eta$ is the fluid viscosity, and $\kappa$ is the thermal diffusivity.
The threshold value for a flat fluid film is about $80$~\cite{Pearson1958}.
The onset of Marangoni convection was also confirmed in~\cite{McDonaldWard2012}
in the particular case of droplets with the contact angle $\theta=90^\circ$.

There are other hydrodynamic effects in droplets and films, which in some
cases play a role in obtaining structured deposits. Solutal Marangoni flow
is due to the influence of surfactants on the surface tension of the
solution. For example, it was shown in~\cite{Marin2019} that this flow can
generate ring-shaped stains in the case of a drop of salt solution.
Finger-like instability in films and droplets can occur when surfactants are
present~\cite{Cachile200259}. The thermocapillary rupture in a film flowing
down a tilted plate was observed in the local heating point~\cite{Zaitsev2007174}.
The Leidenfrost effect is another example of a classic effect in droplet--film
systems that is actively used and studied and can affect a pattern formation~\cite{Leidenfrost1756,Kabov201760,Kruse2015,Zaitsev2017094503}.

The heterogeneous heat and mass transfer in an evaporating liquid is an
essential condition for the emergence of structured deposits not only in
the case of the coffee-ring effect but also in many other problem
statements. In liquid sessile droplets, such processes naturally occur
in the vicinity of the contact line. Heterogeneities in droplets and
liquid films can also be induced by producing ambient conditions
described below. The ability to create structures in a controlled manner
using such external influences is the fundamental in the evaporative
lithography method.

Progress in understanding the coffee-ring effect and other physical
processes occurring in evaporating droplets and films, substantial
developments in experimental research, and the wide interest in these
phenomena in many applications led to the significant activity of
experimenters and theorists in this field over the last twenty years.
Review papers cover many details of this subject: dynamics and evolution
of thin liquid films~\cite{CrasterMatar2009}, spreading of surfactants in
thin liquid films~\cite{Afsar-Siddiqui2003}, structural evolution of drying
droplets of biological liquids~\cite{Tarasevich2004,Yakhno2009},
condensation and coalescence of droplets~\cite{Sikarwar2012}, wetting and
spreading effects~\cite{Bonn2009,Edalatpour2018,Bhushan2011}, organized
structures of particles on a substrate~\cite{Han20121534,Sefiane2018,Patil2019},
inkjet-printed photonic crystal structures~\cite{Wang20136048}, evaporation
of thin films of colloidal solutions~\cite{Routh2013}, self-assembly of 3D
structures of nano- and microparticles~\cite{Brugarolas2013}, hydrothermal
waves in evaporating sessile droplets and the effect of substrate
properties~\cite{Kovalchuk2014}, mass transport in evaporating
droplets~\cite{Hu2012,Zang2019}, stain geometry during evaporation of a
surfactant solution~\cite{Erbil2015}, evaporation of nanoparticle-laden
droplets~\cite{Zhong201513}, control of Marangoni convection in liquid
films~\cite{Gambaryan-Roisman2015319}, hydrodynamics in thin liquid
films~\cite{Karakashev2015}, properties and applications of liquid marbles
(droplets wrapped by micro- and nanoparticles)~\cite{Bormashenko2016}, heat
and mass transfer near the liquid--gas interface on heated
substrates~\cite{Ajaev2017918}, methods for self-assembly of monolayers of
colloidal particles~\cite{Lotito2017217}, mitigation, suppression, and use
of the coffee-ring effect~\cite{Mampallil201838,Zhao201912029}, wetting and
evaporation effects~\cite{Brutin2018}, deposits appearing on a substrate
after drying of multicomponent droplets~\cite{Tarafdar2018}, hydromechanical
and mechanical instabilities during drying of solution
droplets~\cite{Pauchard201832}, heat transfer associated with the
liquid--vapor phase change on functional micro- and nanostructured
surfaces~\cite{Wen20182307}, stratification of particles upon drying
of colloidal films~\cite{Schulz20186181}, multiscale assembly of organic
electronics~\cite{Patel2018}, simulation of spreading of surfactant-laden
droplets~\cite{Theodorakis2019}, and the role of surfactants in obtaining
structures on substrates~\cite{Shao2020}. The self-assembly of organized
structures during evaporation was reviewed in
monographs~\cite{Lin2012,Innocenzi2013}, evaporation and wetting, in
books~\cite{Sazhin2014,Brutin2015}, and cracks occurring during drying
and associated structures, in~\cite{Goehring2015}.

\begin{figure}
\centering
  \includegraphics[width=0.85\linewidth]{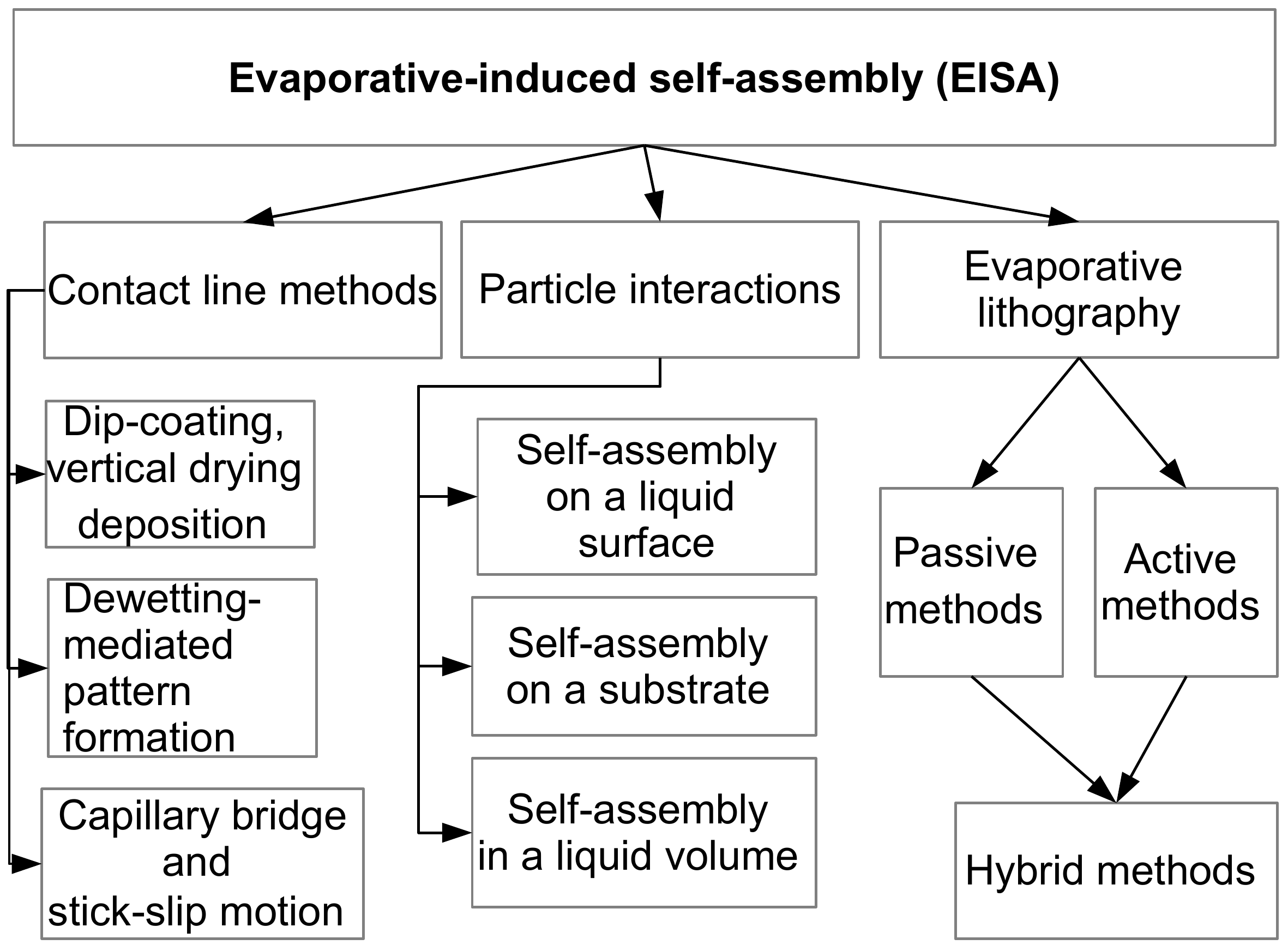}\\
  \caption{
  EISA methods classification.
  }
  \label{fig:EISA_classification}
\end{figure}

Many methods for obtaining structured deposits based on the self-assembly
of micro- and nanoparticles as a result of evaporation have been
proposed~\cite{Dommelen201897}. The studies in this direction are known as
``evaporative-induced self-assembly'' (EISA). For example, EISA is actively
used to produce porous mesostructures~\cite{Soler-Illia2002,Brinker2004,Grosso2004,Bormashenko2005}.
Here, we propose the following classification of EISA methods
(Fig.~\ref{fig:EISA_classification}). Of course, this division into
subgroups is very tentative, and some approaches can often simultaneously
combine principles that are characteristic of different subgroups. The
first subgroup of methods is based on using contact line processes
(``contact line methods'' in Fig.~\ref{fig:EISA_classification}). For
example, a plate is placed above the film. It can be flat~\cite{Li2014,Jabal2018}.
Sometimes, the plate is characterized by a varying curvature due to
swelling~\cite{Li2018124}. In other cases, its geometry is a spherical
segment~\cite{Xu2006066104,Xu2007}, a semicylinder form~\cite{Li2013}, an
ad hoc geometric shape~\cite{Hong2009512}, or a template with groove/ridge
surface topography~\cite{LiS2018}. Normally, the substrate is immobile
throughout the process. For example, in the experiment~\cite{Li20181850360},
the substrate was heated to 130--140\,$^\circ$C to obtain a conductive and
transparent structured thin film of nanotubes used in flexible electronics.
The creation of periodic nanotube stripes can be controlled with
surfactants~\cite{LiH2014}. In another experiment~\cite{He201716045}, the
substrate was slowly displaced relative to the plate located at a slight
angle. This allowed obtaining a prolonged film of perovskite. Such
photoactive layers can be used as flat solar cells. A periodic deposition
pattern of a binary colloid mixture can be also obtained under a moving
plate~\cite{Das2018}. This method was improved in~\cite{Choudhary2018} by
using an oscillating blade and a bending actuator. This provides an
additional possibility to manage the pattern being formed. One more key
parameter of this technique is the oscillation frequency of the blade tip.
The mentioned EISA methods are normally based on the capillary bridge and
stick--slip motion effect~\cite{Xu2006066104} (see Fig.~\ref{fig:EISA_classification}).
Essentially, it includes wetting two surfaces by a film in the gap between
them~\cite{Xu2006066104} and mass transfer with a capillary flow toward
the contact line where intensive evaporation occurs. Sometimes, the contact
line moves periodically, which generates concentric
deposits~\cite{Maheshwari2008,Bodiguel2009,Bodiguel2010}. This also applies
to the drying of a liquid film with nanoparticles on the surface of a layer
of another liquid~\cite{Dong2011}. Sometimes, fingering instabilities
produce spoke patterns~\cite{Xu2007}. The authors of~\cite{Li2013}
experimented with DNA nanowires and concluded that spoke patterns can be
obtained instead of concentric rings by changing the pH of the solution.
The drying deposition method on a vertical plate works by the same
principle~\cite{Nakamura200433,Lozano20079933,Cao2010} (see
Fig.~\ref{fig:EISA_classification}). A meniscus is formed near the plate
as a result of liquid wetting. The capillary flow transfers particles into
this area. As the fluid dries, a particle layer or a periodic pattern is
thus formed on the plate. If the plate is withdrawn with some velocity,
this is the dip-coating method~\cite{Zhang2008,Li2011,Grosso2011}. For
example, the authors of~\cite{Jang2012} used dip-coating to produce organic
transistors. We also mention the Langmuir--Blodgett method, which is close
to dip-coating. This technique transfers a layer of particles from the
liquid surface to a solid surface~\cite{Lenhert2004,Chen2006}. The formation
of micro- and nanoparticle structures often occurs in areas of film rupture
and as the dry patch expands further~\cite{Thiele2009,Stannard2011,Li2017}.
This method is known as dewetting-mediated pattern formation, which is also
included in the subgroup of contact line methods
(Fig.~\ref{fig:EISA_classification}). In review~\cite{Thiele2014}, both
passive and active experimental setups for controlling the contact line
motion were described. Studying the three-phase boundary dynamics in such
methods is essential because it has a significant impact on the sediment
structure. We note that substantial progress has been achieved in studying
the evaporation flux structure in the contact line
area~\cite{Morris2000,Morris2001,Morris2014} and its effect on the
formation of the wetting angle for evaporating sessile droplets with
good wettability~\cite{Rednikov2013,Rednikov2017,Rednikov2019}.

Another subgroup of methods is based on particle interaction forces
(Fig.~\ref{fig:EISA_classification}). These include electrostatic and
intermolecular interaction forces (van der Waals
forces)~\cite{Warner2003,Liang2007,Kovalchuk2019,Molchanov2019}. We also
mention the capillary particle interaction~\cite{Denkov1993,Kralchevsky1994,Park2011}.
Particles assemble directly on the
substrate~\cite{Denkov1992,Liu2002,Nakamura200433}, on the free
liquid surface~\cite{Bigioni2006,Park2011,LiW2019}, or in the
bulk~\cite{Qiu2010,Wang2012,Wang2014}.

Some methods are difficult to assign to EISA, but they also allow forming
certain patterns. Silver nanowires were used in~\cite{Lim2017} to form
stretchable electrodes with excellent electromechanical stability
subjected to periodic mechanical deformation. An array of micron-sized
droplets was applied by spray coating to the surface. The nanowires inside
the microdroplets were bent into curved shapes by the elasto-capillary
interaction while the microdroplets were moving toward the solid surface.
After the droplets impacted on the substrate, rings were formed from
overlapping elastic nanowires. Such structures can well tolerate bending
and stretching while retaining their electrical conductive
properties~\cite{Lim2017}.

The above methods of forming structured coatings are difficult to control,
especially in real-time. The precise pattern except its qualitative
properties is hard to predict. Although work is also progressing in this
direction~\cite{He201716045,Choudhary2018,LiS2018}, another logical
continuation of the development of these methods is the emergence of
a narrower direction called ``evaporative lithography''
(Fig.~\ref{fig:EISA_classification}). According to the theory of Deegan et
al., which explained the coffee-ring effect, it is possible to control the
particle sedimentation process by manipulating the vapor concentration near
the two-phase boundary~\cite{Deegan1997,Deegan2000,Deegan2000475}. The
method called ``evaporative lithography''~\cite{RouthRussel1998,Harris2007}
was developed based on this theory. The essence of this method is to create
ambient conditions for nonuniform evaporation from the colloidal liquid
surface.

There are other lithographic approaches to creating polymer coatings with
topographic patterns at the micro- and nanoscales. Structured colloidal
assemblies are sometimes created on electrically~\cite{Aizenberg2000,Zheng2002569}
and topographically modified substrate
surfaces~\cite{Blaaderen1997321,Lin20001770,Lee20045262,Malainou201313743}.
But such approaches are not easy to apply to other classes of soft materials.
In nanoimprint lithography, the template is in contact with the surface
under pressure~\cite{Yasuda2010}. This sometimes leads to defects in the
template itself. Such defects affect the shape of the resulting patterns.
Evaporative lithography implies no mechanical impact on a template.
Therefore, its integrity is preserved in the process. The method of
photolithography is widely known, but this multistage process requires
special equipment and materials~\cite{Moreau1988}. Structured surfaces of
polymers can be obtained by different molding methods, such as nanoscale
injection molding~\cite{Kim2010}, capillary molding~\cite{Kim1995581}, and
ultraviolet nanoimprinting~\cite{Jeon2008}. These methods have a number of
disadvantages, such as lack of process flexibility and the need for
expensive equipment. Nanosphere lithography (colloidal bead self-assembly)
allows obtaining regular structures only of a single spatial shape when
nanoparticles are deposited in the cavities between densely packed
microspheres~\cite{Li2009,Li2009969,Zhang2010,Pi20133372}.
Microstereolithography requires using special materials
(photopolymers)~\cite{Wu2002,Wu2006}. Moreover, there is often a problem of
the destruction of the printed 3D microstructure under the influence of
capillary forces (the stiction problem)~\cite{Wu2002,Wu2006}.

The EISA methods and, in particular, evaporative lithography are usually
one-step methods. They do not require expensive equipment and materials.
Moreover, they can be applied to almost any surfaces. Evaporative
lithography, compared with other EISA methods, is a more flexible approach
that allows considerable variation in the resulting patterns. In some cases,
patterns are more stable (e.g., when particles are thermally or chemically
sintered). Such methods are suitable for various applications in optical and
microelectronics, health care, nanotechnologies, and other fields (see
Sec.~\ref{applications}). Evaporative lithography therefore attracts
increasing attention. Convection is caused by the nonuniform evaporation of
colloidal film and transfers particles to the intense evaporation area. The
vapor flow density can be controlled along the free surface of the liquid
layer, for example, by placing a mask above the droplet~\cite{Harris2007}.
The obtained structures replicate the shape of the mask holes, which play
the role of templates. A template can be not only a solid obstacle but also
many other things (see Secs.~\ref{sec:experiments} and~\ref{sec:models}
below). The method of evaporative lithography has two restrictions. First,
the total evaporation time increases significantly because the mask blocks
most of the liquid surface. Second, because glassy polymer particles are
incapable of forming a solid film, only brittle coatings with cracks are
obtained for polymer melting temperatures above room temperature without
additional efforts. This problem is solved by introducing an IR source to
the system~\cite{Routh2011}. Local heating creates intensive evaporation
under holes in the mask. The particles heated by infrared light sinter well,
and there are hence no cracks in the resulting structure. This method is
simpler and cheaper and can be applied to almost any substrate without
pretreatment compared with other methods of creating textured layers. For
example, evaporative lithography was used experimentally~\cite{Zhao2016} to
demonstrate the possibility of creating materials with localized functions
such as slipperiness and self-healing.

The purpose of this review is to present the existing achievements in
evaporative lithography and to analyze how this method can be further
developed. Section~\ref{sec:experiments} deals with the main experimental
works in this field. It describes passive, active, and hybrid methods of
controlling the formation of colloidal particle structures
(Fig.~\ref{fig:EISA_classification}). It also provides examples of the
practical use of these methods and discusses dimensionless parameters.
Mathematical models of this class of phenomena are discussed in
Sec.~\ref{sec:models}. The summary (Sec.~\ref{sec:conclusion}) provides
conclusions and discusses possible perspectives for the development of
evaporative lithography.

\section{Experimental achievements in evaporative lithography}
\label{sec:experiments}

\subsection{Passive methods for controlling particle sedimentation}
\label{sub:PassiveMethods}

Existing evaporative lithography methods can be provisionally divided into
passive and active methods. The difference is that active methods allow for
real-time control of the pattern structure being formed, while passive
methods do not provide such a possibility. The control in passive methods is
achieved by adjusting key static parameters before starting the process.
Examples of such parameters include the distance between the mask and
liquid~\cite{Harris2007}, the initial solution concentration~\cite{HarrisLewis2008},
the size of mask holes~\cite{Harris2007}, the substrate
material~\cite{Georgiadis2013} or tilt angle~\cite{Kimura2003}, the distance
between two adjacent droplets~\cite{ChenL2009,Pradhan2015,Hegde2018,Hegde2019348},
the substrate temperature gradient (in case of nonuniform
heating)~\cite{Malla2019}, etc. In this section, we focus on describing
experiments related to passive methods in evaporative lithography.

\begin{figure}[htb!]
\centering
  \includegraphics[width=0.85\linewidth]{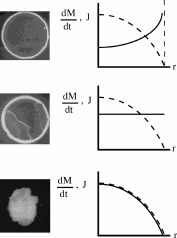}\\
  \caption{Types of deposits (left) and the corresponding types of the
  function $J$ (solid line) relative to the droplet profile $h$ (dashed
  line). Reproduced with permission from Ref.~\cite{Deegan2000} (Copyright
  2000 by The American Physical Society).}
  \label{fig:DeeganExperiment}
\end{figure}

The experiment of Deegan et al.~\cite{Deegan2000} showed a ring-shaped
pattern (Fig.~\ref{fig:DeeganExperiment}) formed as a result of intensive
evaporation near the liquid--substrate--air boundary. The position of this
ring corresponds to the location of the three-phase boundary (the pinned
contact line). An analytic equation was derived in~\cite{Deegan2000475} to
describe the ring width $w(t)$ as a function of time,
\begin{equation*}\label{eq:RingWidth}
w= R\sqrt{\frac{\phi}{4p}} \left[ 1- \left( 1-
\frac{t}{t_\mathrm{max}} \right)^\frac{3}{4} \right]^{\frac{2}{3}},
\end{equation*}
where $\phi$ is the volume fraction of colloidal particles, $p$ is the
packing density factor, and $t$ and $t_{\max}$ are the respective time and
full time of evaporation. Under ambient conditions, the vapor flux density
$J(r,t)$ normally increases from the center ($r=0$) to the droplet periphery
($r=R$), whose height is much less than the base radius
($h(r=0,t)\ll R$)~\cite{Deegan1997}. The droplet base refers to the boundary
where the liquid contacts the substrate. If the vapor flux density is
uniform along the free surface of the droplet, a ring pattern is also formed
(Fig.~\ref{fig:DeeganExperiment}). Such a spatial function $J$ can be
obtained if the ambient humidity near the free droplet surface is relatively
high. For this, the authors of~\cite{Deegan2000} placed a substrate with a
droplet on the water surface in a small container. In these two experiments,
compensatory flows appeared that transferred the suspended matter to the
contact line.

In the third experiment~\cite{Deegan2000}, the droplet was covered with a
cap with a hole in the middle. As a result of intensive evaporation near
the hole, compensatory flow carried the colloids to the central area. After
the droplet dried, a spot-shaped deposit was observed on the substrate
(Fig.~\ref{fig:DeeganExperiment}). Based on the theoretical and experimental
results, the authors of~\cite{Deegan2000} formulated the idea of possible
control of the deposition process by manipulating the vapor concentration
around the droplet.

\begin{figure}[htb!]
\centering
  \includegraphics[width=0.85\linewidth]{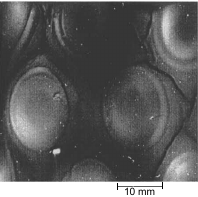}\\
  \caption{Film with selective evaporation in circular regions. Reprinted
  with permission from Ref.~\cite{RouthRussel1998} (Copyright 1998, American
  Institute of Chemical Engineers (AIChE)).}
  \label{fig:RouthRussel_exper98}
\end{figure}

Another experimental team~\cite{RouthRussel1998} performed measurements with
a latex film drying on a substrate. The initial thickness of the liquid
layer was 0.15 mm. In some places, evaporation was blocked by placing a
coating with regularly spaced holes. The distance between the centers of
adjacent holes was 16 mm. Such a coating was located in close proximity to
the free surface of the film without the coating and film touching directly,
but the gap between them was sufficiently small. This allowed evaporation to
be blocked in certain local areas of the film.
Figure~\ref{fig:RouthRussel_exper98} shows the structure of the latex film
dried using this method~\cite{RouthRussel1998}.

The results of profilometry~\cite{RouthRussel1998} showed that the maximum
thickness of the final film is about 90 $\mu$m in areas where evaporation
occurred and about three times less in places where evaporation was
suppressed.

The evaporation rate can be adjusted by selecting the liquid. For example,
in~\cite{Rieger2003}, ethylene glycol is used, which evaporates relatively
slowly. As a result, the capillary flow rate is less than 1 $\mu$m/s. The
flow slowly transfers microspheres to the wall of an open cylindrical cell.
Nanoliter cells represent interest for labs-on-a-chip~\cite{Rieger2003}.

The term ``evaporative lithography'' was suggested in~\cite{Harris2007}. A
mask with holes placed above the film surface was used in the experiment.
The authors of~\cite{Harris2007} considered key parameters such as the
initial volume fraction of the solution $\phi_\mu$, the mask design
(geometric parameters), and the gap size between the film and the mask
$h_g$. Masks contained a hexagonal array of holes of different diameters
$d_h$ with a pitch of $P$ (distance between centers of adjacent holes). The
resulting pattern consisted of individual particle aggregates several
microns high. Their cross-sectional size was approximately equal to the
value of $d_h$. Unary and binary colloidal films were investigated
in~\cite{Harris2007}. Silica microspheres and polystyrene nanoparticles were
taken. The authors of~\cite{Harris2007} estimated the microsphere
sedimentation time by dividing the initial height of the film $h_i$ by the
Stokes velocity $U_0=2a^2\Delta\rho g/(9\mu)$, where $a$ is the particle
radius, $\Delta\rho$ is the particle and water density difference, $g$ is
the acceleration of gravity, and $\mu$ is the dynamic viscosity of the
liquid. The full evaporation time ($\approx120$ min) significantly exceeds
the sedimentation time ($\approx2$ min). But in the case of a dilute
solution, almost all particles migrated to the areas under holes. In our
opinion, this is possible with weak adhesion of particles to the substrate
that does not prevent their drift in the convection direction.

\begin{figure}[htb!]
\centering
  \includegraphics[width=0.95\linewidth]{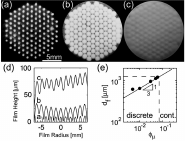}\\
  \caption{Optical images of the films obtained from colloidal suspensions
  with $\phi_\mu=$ (a)~0.005, (b)~0.1, (c)~0.3; (d)~the corresponding height
  profiles; (e)~the dependence of the element diameter $d_f$ on the initial
  volume fraction $\phi_\mu$ of the solution. Reproduced with permission
  from Ref.~\cite{Harris2007} (Copyright 2007 by The American Physical
  Society).
}
  \label{fig:patterVsVolFrac}
\end{figure}

The influence of the initial volume fraction of the solution was studied at
fixed values of the parameters $d_h$, $P$, and $h_g$~\cite{Harris2007}. The
value of $\phi_\mu$ varied in the range from 0.005 to 0.3.
Figure~\ref{fig:patterVsVolFrac} shows optical images of the resulting
deposits. Their patterns differ for different values of $\phi_\mu$. Discrete
elements of sedimentation are formed under the holes when the initial volume
fraction of particles is small (Fig.~\ref{fig:patterVsVolFrac}a). Their
diameter $d_f$ is relatively small, $d_f<P$. For $\phi_\mu>\phi_\mu^*$, the
diameter of such elements $d_f$ corresponds to the distance $P$, $d_f=P$. In
such cases, a relatively thick continuous film with a grooved surface is
formed (Figs.~\ref{fig:patterVsVolFrac}b and \ref{fig:patterVsVolFrac}c).
Dome-shaped prominences can be seen under the holes
(Fig.~\ref{fig:patterVsVolFrac}d). The critical initial volume fraction
$\phi_\mu^*$ of microparticles was defined in~\cite{Harris2007} based on
geometric considerations, $\phi_\mu^*\approx\pi\phi_fP/200\sqrt{3}h_i\approx0.07$,
and experimentally (Fig.~\ref{fig:patterVsVolFrac}e). The volume fraction of
randomly packed microspheres is $\phi_f\approx0.64$.

The dependence of the dried film height difference $\Delta h=h_{\max}-h_{\min}$
on the parameter $h_g/P$ was measured at a fixed $\phi_\mu$ for different
$d_h$, $P$, and $h_g$~\cite{Harris2007}. The value of $\Delta h$ is small
for $h_g/P\geq0.3$, which evidences of the poor segregation of particles. As
$P$ increases with a fixed value of $h_g/ P$, the value of $\Delta h$
increases.

\begin{figure}[htb!]
\centering
  \includegraphics[width=0.95\linewidth]{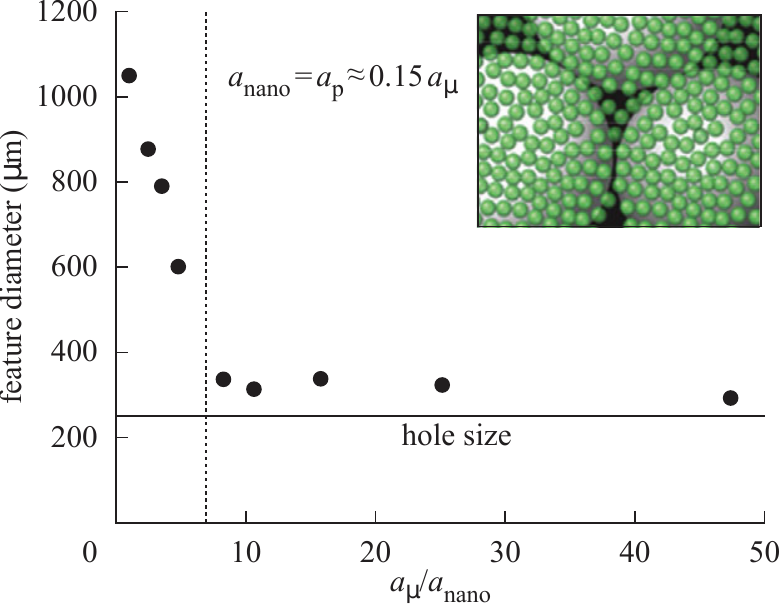}\\
  \caption{The diameter of the element (sedimentation of nanoparticles under
  the mask hole) as a function of $a_\mu/a_\mathrm{nano}$ (the inset
  schematically shows filling of pores between microspheres with
  nanoparticles for a high ratio of sizes). Reproduced from
  Ref.~\cite{Harris2009} with permission of the Royal Society.}
  \label{fig:binarryHarris}
\end{figure}

The process is somewhat different in the case of a two-component solution
with volume fractions $\phi_\mu=0.3$ of microspheres and $\phi_n=10^{-3}$ of
nanoparticles~\cite{Harris2007}. First, a continuous relief film is obtained
from the microspheres. Second, nanoparticles assemble into discrete clusters
under the holes in the mask. The porous medium formed by the presence of
cavities between microspheres provides an additional driving force to the
nanoparticles. The pore radius is $a_p\approx0.15a_\mu$, where $a_\mu$ is
the microsphere radius. Liquid residues in pores under the open area form
menisci with capillary pressure induced at their surfaces. The pressure
difference accelerates the transfer of nanoparticles to the places below the
holes. Nanoparticles migrate into a network of cavities when the radius of
the nanoparticles is $a_\mathrm{nano}\ll a_p$.  The dependence of the
resulting particle distribution structure on the size ratio of micro- and
nanoparticles in a binary system was studied separately in~\cite{Harris2009}.
The plot (Fig.~\ref{fig:binarryHarris}) shows that for $a_\mu/a_\mathrm{nano}
\geq7$, the element diameter is close to the hole size ($d_h=250\;\mu$m).
The value~7 corresponds to the case where the size of nanoparticles is equal
to the pore size.

\begin{figure}[htb!]
\centering
  \includegraphics[width=0.85\linewidth]{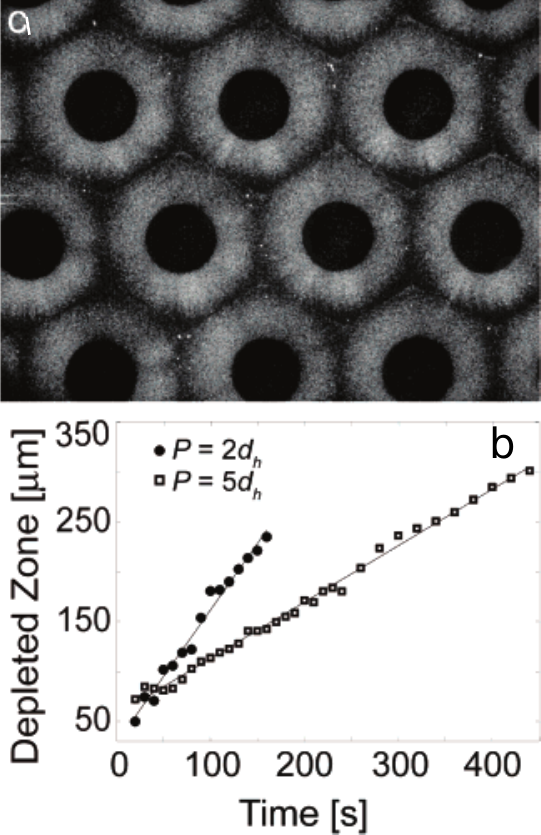}\\
  \caption{(a) Fluorescent image of the drying film ($\phi_i$=0.001) at
  190~s after placing the mask ($d_h=250\;\mu$m, $P=2d_h$); (b)~an increase
  in the size of the depleted zone under the hole over time. Reprinted with
  permission from Ref.~\cite{HarrisLewis2008} (Copyright 2008 American
  Chemical Society).
}
  \label{fig:inversedPattern}
\end{figure}

The possibility of obtaining an inverted pattern from deposited particles
was demonstrated in~\cite{HarrisLewis2008}. Under certain conditions,
particles accumulate under covered areas of the mask. A sort of inverted
pattern is thus obtained relative to the pattern of holes in the mask.
Silica microspheres with a radius of $492\pm58$~nm suspended in ethanol were
used in~\cite{HarrisLewis2008}. The particles were stabilized in solution
using a surfactant. The initial volume fraction $\phi_i$ of the solution was
varied in the range from 0.001 to 0.15. Fluorescence microscopy and tracers
(a small fraction of particles was marked with fluorescent) were used in the
experiment for direct observation and visualization of flows (see the
Supplemental Material video~\cite{HarrisLewis2008}). Two types of masks were
used: hexagonal arrays of circular holes and parallel strips. When the
initial concentration of the solution is low, particles accumulate in the
masked areas (Fig.~\ref{fig:inversedPattern}a), where evaporation is slow.
At the same time, a depleted zone appears under the holes and grows linearly
with time (Fig.~\ref{fig:inversedPattern}b). The fact is that with intense
evaporation under the mask holes, the local temperature of the film surface
decreases. The variation of temperature and the associated surface tension
gradient lead to the Marangoni convection. This convection flow is directed
from a low surface tension area to a high surface tension area where the
film surface temperature is relatively low. The typical particle velocity
in such a flow is about $10^{-3}$ m/s. For comparison, the velocity of the
compensatory flow caused by evaporation is $LE/h_i\approx5\times10^{-7}$ m/s.
Here, $L=2.5\times 10^{-4}$ m is the typical size of the convection flow
cell, the evaporation rate is $E=2.2\times10^{-7}$ m/s, and the initial film
height is $h_i=10^{-4}$ m. Disappearance of the Marangoni effect was
observed~\cite{HarrisLewis2008} when the solution reached a critical volume
fraction $\phi_c\approx0.22$. When the values of $\phi_i$ are relatively
high, structure inversion is not observed, because the system is mostly
affected by the compensatory flow.

\begin{figure}[htb!]
\centering
  \includegraphics[width=0.85\linewidth]{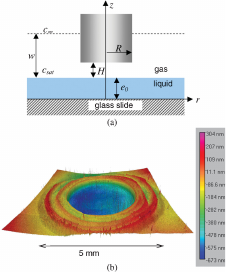}\\
  \caption{(a) Setup for film evaporation control: the film is schematically
  shown at the initial moment. (b)~A dip with a rim is formed under the
  obstacle: the profile of the dried layer obtained with a film thickness
  $e_0=3.5\;\mu$m, rod diameter $2R=5$ mm, and a gap $H=1$ mm observed using
  an optical profilometer (zero of the vertical axis corresponds to the film
  level away from the obstacle). Reproduced with permission from
  Ref.~\cite{Parneix2010} (Copyright 2010 by The American Physical Society).
}
  \label{fig:DipsAndRims}
\end{figure}

In~\cite{Parneix2010}, a rod was used as an obstacle locally blocking
evaporation above the thin film surface (Fig.~\ref{fig:DipsAndRims}a).
Nonuniform evaporation resulted in the appearance of a dip under the
obstacle surrounded by a rim (Fig.~\ref{fig:DipsAndRims}b). The diameter of
the rim corresponds to the diameter of the cylindrical rod. The experiment
used a glass substrate and colloidal silica particles with a radius of about
11 nm. The rod radius $R$ and the gap $H$ between the rod and the film were
varied in the experiment~\cite{Parneix2010}. These two key parameters allow
adjusting the width and depth of the dip. The liquid--air interface is
curved in the process of nonuniform evaporation~\cite{Parneix2010}. It
becomes convex under the obstacle. The solid dried layer starts forming at a
distance from the rod location. The solid--liquid phase boundary gradually
propagates toward the rod. The liquid flows from the closed area to the open
area where evaporation is intense. In addition, the liquid is pumped out of
the closed area by filtration. The resulting porous structure of the solid
layer sucks the liquid into the capillaries because of the pressure gradient.
The area of film under the rod dries last.

\begin{figure}[htb!]
\centering
  \includegraphics[width=0.99\linewidth]{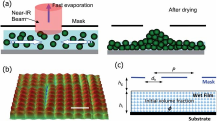}\\
  \caption{(a) A film evaporating under a mask illuminated by an IR lamp;
  (b)~the resulting pattern (scale bar is 3 mm); (c)~physical parameters of
  the system. Reproduced from Ref.~\cite{Routh2011} with permission from The
  Royal Society of Chemistry.
}
  \label{fig:IRAEL}
\end{figure}

IRAEL (infrared radiation-assisted evaporative lithography~\cite{Routh2011})
is a modification of the approach proposed in~\cite{Harris2007} with a
mask. To increase the evaporation rate, an infrared lamp is added to the
system (Fig.~\ref{fig:IRAEL})~\cite{Routh2011,Georgiadis2013}. If the
melting point of polymer particles exceeds room temperature, then external
IR heating enables the coalescence of particles. A relief glass film
without cracks is formed after drying. For comparison, structures of solid
polymer particles obtained using the standard method~\cite{Harris2007} are
brittle. The thickness of thin solid films with a stable periodic structure
can vary from several microns to submillimeter sizes~\cite{Routh2011,Georgiadis2013}.
In~\cite{Harris2007}, the evaporation process lasted about two hours because
the mask blocks a part of the free film surface. In the case of IR heating,
the process lasts several minutes or dozens of minutes depending on the lamp
power~\cite{Routh2011,Georgiadis2013}. High-power lamps should be used in a
way that avoids overheating the system. The issue is that a boiling liquid
adversely affects the formation of the required pattern~\cite{Georgiadis2013}.
The radiation power density is $P_d=P_E/(2\pi r_lL)$~\cite{Routh2011} or
$P_d=P_L/(4\pi r_l^2)$~\cite{Georgiadis2013}, where $r_l$ is the distance
from the film surface to a light source, $L$ is the length of a cylindrical
emitter, $P_E$ and $P_L$ are the respective emitter and lamp powers.

The authors of~\cite{Routh2011} used aqueous latex films and an IR lamp
($P_L=250$ W) in their experiment to study the influence of $h_i$, $\phi_i$,
$h_g/P$, and $P$ on $\Delta h$. The value of $\Delta h$ linearly increases
as the initial film height $h_i$ increases. The vertical distance between
the peak and the valley $\Delta h$ also increases with pitch $P$. This only
occurs until a certain value ($P\approx 3$~mm) is achieved. Further increase
in $P$ does not affect the increase in $\Delta h$. An increase in the
initial volume fraction $\phi_i$ causes an increase in $\Delta h$ if
$\phi_i<\phi_c$. The value of $\Delta h$ decreases if the initial volume
fraction is above a critical value of $\phi_c\approx0.35$. A decrease of
$\Delta h$ is also observed when $h_g/P$ is increased.

Marangoni flow in IRAEL is expected to be directed from open areas to those
covered with a mask having relatively cold temperatures, but the
authors~\cite{Routh2011} observed an accumulation of substance in the open
areas. This indicates that the compensatory flow resulting from evaporation
dominates. For a theoretical estimate, they used a characteristic
thermocapillary flow velocity $V_M=(h_i/L)(d\gamma/dT)(\Delta T/\mu)$ and a
typical compensatory flow velocity $V_E=LE/h_i$, where the capillary length
is $L=h_i\left(\gamma/(3\mu E)\right)^{1/4}$, $\gamma$ is the surface
tension, $\mu$ is the solution viscosity, $E$ is the evaporation rate, and
$\Delta T$ is the temperature gradient. With a high ratio of $\psi=V_M/V_E$,
Marangoni flow prevails. The compensatory flow prevails when $\psi$ is low.
As the volatile substance is lost, the solution concentration $\phi$ grows along
with the viscosity $\mu(\phi)$. As $\mu$ increases, $\psi$ decreases. This
estimate thus confirms the observation of the fact that Marangoni flow
disappears with $\phi\geq0.22$~\cite{HarrisLewis2008}.

The effect of IR emitter power on the system was studied
in~\cite{Georgiadis2013}. Carbon IR emitters with a power value from
1,600~W to 2,400~W were used in the experiment~\cite{Georgiadis2013}. The
radiation power density $P_d$ was measured as a function of $r_l$ and $P_E$.
The value of $P_d$ increases nonlinearly as the emitter power increases and
the distance between the IR light source and the film decreases. With a
higher power $P_E$, the temperature in the film rises faster over time. This
allows adjusting the evaporation rate. Different substrate materials (brass,
copper, steel, glass, paper, aluminum alloy, polyester) were used
in~\cite{Georgiadis2013}. The experiment showed that the type of substrate
material affects a parameter of the formed structure such as $\Delta h$. The
difference in $\Delta h$ is most likely due to the influence of heat
conductivity, absorption of IR radiation by the substrate, and the surface
energy of the material. To explain the empirical data obtained, more
detailed research and mathematical modeling are needed.

\begin{figure}[htb!]
\centering
  \includegraphics[width=0.99\linewidth]{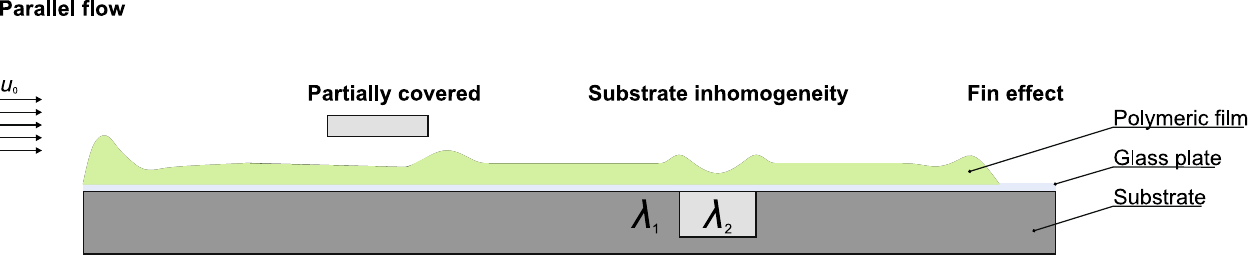}\\
  \caption{Possible causes of curvature of the liquid film free surface.
Adopted from Ref.~\cite{Cavadini2015} with the permission of Springer.
}
  \label{fig:surfaceCurvatureReasons}
\end{figure}

There are several reasons for the deformation of the free surface of the
liquid film, which plays an important role in evaporative lithography:
surface disturbance by airflow, the edge effect in the three-phase boundary,
partial covering of the film area with a plate, evaporation from an
inhomogeneous substrate, etc. (Fig.~\ref{fig:surfaceCurvatureReasons}). One
of the promising directions is the use of composite substrates with variable
heat properties~\cite{Cavadini201324,Cavadini2015}.

An aluminum substrate with a Teflon strip applied was used
in~\cite{Cavadini201324,Cavadini2015}. The respective heat conductivity
coefficients of aluminum and Teflon are $\lambda_1$ and $\lambda_2$ in
Fig.~\ref{fig:surfaceCurvatureReasons}. A thin glass plate with a liquid
film was placed over the substrate. Such an approach allows using the
substrate as a template many times because the pattern is formed on the
glass plate. Teflon emits heat slower than aluminum because the latter has a
relatively high heat conductivity. Therefore, the local part of the liquid
film is relatively cold because of evaporation in the Teflon area. The
aluminum substrate was placed in a drying channel through which air was
blown. The airflow rate $u$ was one of the parameters studied
in~\cite{Cavadini201324}. Moreover, the experiment was conducted at
different temperatures $T$. In the considered parameter ranges
($u=$ 0.5--1.5~m/s and $T=$ 25--40\,$^\circ$C), no significant differences
in the final patterns of deposits were observed. A solution of methanol and
polyvynilacetate was used~\cite{Cavadini201324,Cavadini2015}. The surface
tension of the solution was measured at different temperatures and
concentrations~\cite{Cavadini201324}. The data on the dependence of the
solution viscosity on concentration were also provided~\cite{Cavadini2015}.
At different time stages of evaporation, several types of instability on the
liquid surface were observed~\cite{Cavadini201324}. This is associated with
the influence of the thermal and solutal Marangoni effects alternatively
dominating in different process stages. A quasi-one-dimensional model was
used in~\cite{Cavadini201324} to calculate the liquid mass fraction and the
film surface temperature variation in time. According to the calculations,
the temperature above Teflon is lower than that above aluminum. The mass
fraction of liquid in the solution decreases faster above aluminum over
time. Lateral mass and heat distribution were disregarded~\cite{Cavadini201324}.
We do not discuss the model~\cite{Cavadini201324} in Sec.~\ref{sec:models},
because it is too simple.

\begin{figure}[htb!]
\centering
  \includegraphics[width=0.99\linewidth]{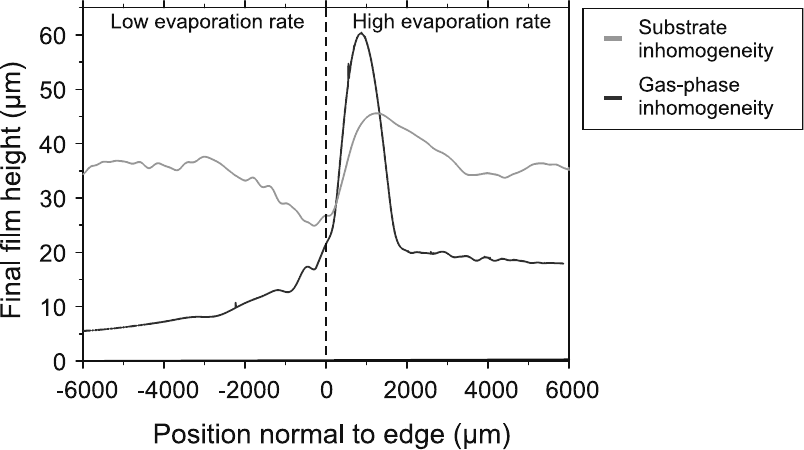}\\
  \caption{Comparison of profiles of the final patterns obtained using two
  methods. Reproduced from Ref.~\cite{Cavadini2015} with the permission of
  Springer.}
  \label{fig:compositeSubstrateVsPartialCovered}
\end{figure}

In~\cite{Cavadini2015}, the method with a composite substrate was compared
with the film evaporation partially covered with a plate on top
(Fig.~\ref{fig:surfaceCurvatureReasons}). While the nonuniform evaporation
in the first case is caused by the heterogeneous heat conductivity of the
substrate, the evaporation in the second method is simply blocked in a
certain area by the plate. Figure~\ref{fig:compositeSubstrateVsPartialCovered}
shows profiles of two final patterns obtained using these two methods. Both
cases demonstrate a deposit peak in the part of the intensive evaporation
area near the boundary of the slow evaporation area. The difference in the
solid film thickness on a composite substrate was $\Delta h\approx20\;\mu$m.
For comparison, in the pattern obtained due to partial blocking of
evaporation with the plate, $\Delta h\approx55\;\mu$m.

The authors paid special attention to visualizing the flow and
measuring its velocity~\cite{Cavadini2015}, focusing on three areas:
(1)~center of the covered area, (2)~the intermediate area, and (3)~the open
area. The maximum flow velocity was measured in the intermediate area,
$v_{\max}\approx$ $98\pm 12$ $\mu$m/s. The observation results showed that
the flow velocity increases initially. After a while, the rate begins to
decelerate because the solution viscosity increases as the solvent is lost.
Details of the measurement method can be found in~\cite{Cavadini2018}.

In some applications, the direction of the compensatory flow should be
reversed to suppress the coffee-ring effect. In~\cite{Yen2018}, the top
central part of a droplet was heated by a CO$_2$ laser. The droplet was
placed on a hydrophobic substrate over a hydrophilic well. The well in the
substrate corresponded to the droplet center. As a result of the heating,
the evaporation rate in the central area was higher than in the periphery.
Consequently, the compensatory flow was directed from the droplet periphery
to its center. Over time, suspended DNA molecules accumulated in the well.
At the same time, the three-phase boundary was displaced (``slip motion'' or
constant contact angle mode). Because the amount of DNA molecules in the
periphery was small, no pinning occurred. After drying, stain-like deposits
were formed. In contrast to the eOMA method~\cite{Anyfantakis2017} (see
section~\ref{subsec:HybridMethods}), the method proposed in~\cite{Yen2018}
does not require using surfactants.

An additional experiment with the heated substrate without using a laser was
performed for comparison~\cite{Yen2018}. It was shown that in this case, a
ring pattern formed in the area of the pinned three-phase boundary. The
boundary was pinned because the DNA molecules accumulated, drifted by the
compensatory flow toward the periphery where the evaporation rate was higher.

In contrast to the results in~\cite{Yen2018}, during point heating of
the droplet center with IR light, particles were carried with the
flow toward the periphery in the experiment in~\cite{Thokchom2014}. And
when the point area at the droplet edge was heated, particles were carried
to the opposite edge. The authors~\cite{Thokchom2014} explained this
transfer by an emerging Marangoni flow. This flow is directed along the
surface from a warm area to a cold area where liquid surface tension is
relatively high. To confirm this, the authors~\cite{Thokchom2014} developed
a mathematical model and performed a series of calculations. The
model~\cite{Thokchom2014} describes buoyancy-induced flow and Marangoni
convection in a constant droplet volume with heating of a surface point. The
model~\cite{Thokchom2014} does not account for the convection--diffusion
equation to describe the particles transfer nor does it account for the
movement of the two-phase boundary or for volume heat flux resulting from
the IR light passing through the liquid. The numerical results are difficult
to attribute to evaporative lithography unlike the experimental part.
Therefore, we do not describe the model~\cite{Thokchom2014} in detail in
Sec.~\ref{sec:models}. For a pure water droplet, an influence of local
heating both on Marangoni flows inside the drop and on the evaporation
kinetics was observed in~\cite{Askounis2017}.

In summary, passive methods in evaporative lithography are usually
based on controlling the geometric characteristics of the
system~\cite{RouthRussel1998,Harris2007,Parneix2010}, the heating
temperature~\cite{Routh2011,Georgiadis2013,Yen2018}, and the physical and
chemical properties of the liquid~\cite{Rieger2003} and the
substrate~\cite{Cavadini201324}. These methods are clear and easy to
implement and can be used in several practical applications (see
Sec.~\ref{applications}).

\subsection{Active methods for controlling particle sedimentation}

Real-time control of particle deposition using active methods is mainly
achieved by regulating key dynamic parameters of the system. As a rule, this
implies some kind of external influence, for example, a powerful light
emitter~\cite{VieyraSalas2012} or an intensive air flow~\cite{Wedershoven2018}.
Parameters of external sources can be controlled during the drying of the
colloidal liquid. In this section, we review the main experimental
achievements regarding active methods in evaporative lithography.

\begin{figure}[htbp]
\centering
  \includegraphics[width=0.95\linewidth]{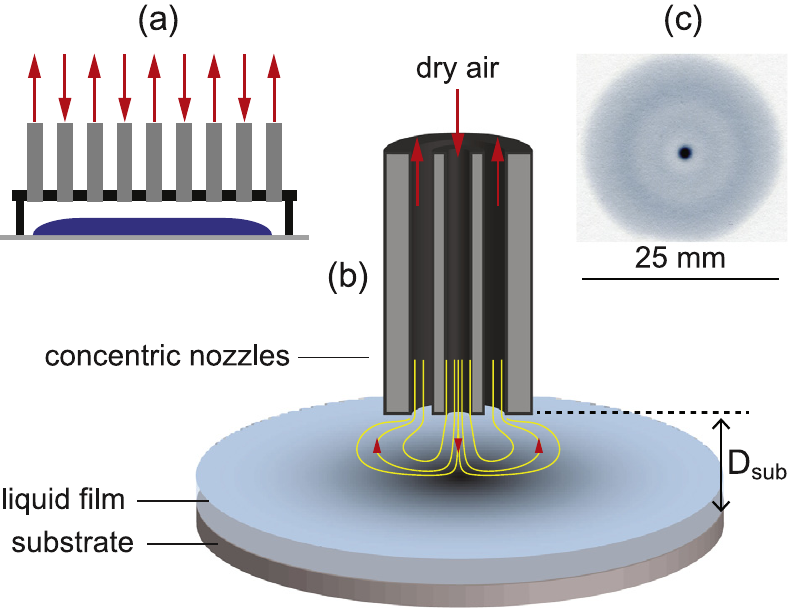}\\
  \caption{ (a) Active control in evaporative lithography schematically.
  (b)~Sketch of the experimental setup. (c)~Image of the resulting pattern.
  Reprinted from Ref.~\cite{Wedershoven2018}. Copyright 2018 with permission
  from Elsevier.
}
  \label{fig:concentricNozzles}
\end{figure}

The vapor concentration above the film surface can be controlled by a dry
air flow. For example, a concentric nozzle was placed above a liquid film
containing nanoparticles. This film was preapplied to a glass
substrate~\cite{Wedershoven2018}. Figure~\ref{fig:concentricNozzles}(b)
shows the sketch of the experimental setup. A nozzle with two holes was used
in~\cite{Wedershoven2018}, but more holes could be made
(Fig. ~\ref{fig:concentricNozzles}(a)). Dry air was blown out through the
inner needle (inner radius $R_1=0.25$~mm, outer radius $R_2=0.42$~mm). Air
together with vapor was pumped in through the outer needle (inner radius
$R_3=0.69$~mm, outer radius $R_4=0.92$~mm). The dynamic parameter in such a
system is the air flow velocity $u$. The maximum velocity of the blown-out
airflow can be expressed as $u_{\max}= 2Q/(\pi R_1^2)$, where $Q$ is the
flow rate. If $Q=1$~mL/min, then we obtain the value $u_{\max}=0.16$~m/s. As
a result of the experiment~\cite{Wedershoven2018}, dry deposition was
obtained under the central part of the nozzle. The spot pattern diameter was
approximately 1~mm (Fig.~\ref{fig:concentricNozzles}(c)). A series of
experiments with different values of $Q$ allowed determining the dependence
of the dry spot diameter and height on the air flow velocity. As $Q$
increases, the diameter decreases, and the height increases~\cite{Wedershoven2018}.
In addition, the authors~\cite{Wedershoven2018} also developed a
mathematical model of this experiment (see Sec.~\ref{sec:models}).

\begin{figure}
\centering
  \includegraphics[width=0.95\linewidth]{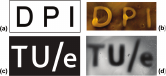}\\
  \caption{(a, c) IR light templates and (b, d) corresponding modulated
  deposition patterns for (b) the dried solution of mesitylene and the red
  light-emitting polymer (pattern width is 2 cm)   and (d) PEDOT:PSS in
  water (pattern width is 2.5 cm). Reprinted with permission from
  Ref.~\cite{VieyraSalas2012}. Copyright 2012 American Chemical Society.
}.
  \label{fig:IR_laser_pattern}
\end{figure}

Infrared (IR) beams can be projected to the solution surface without a mask.
For this, an IR laser and a digital multimirror device were used
in~\cite{VieyraSalas2012}. The authors presented the proposed technology as
an active method for controlling the deposition of the substance during
evaporation, suitable for use in the production of organic electronic materials.
Such a system allows engineers to realize both time-based modulation of
illumination intensity and spatial modulation very rapidly in a parallel
manner. Two different types of solutions were used in the experiment
in~\cite{VieyraSalas2012}. (1)~An aqueous dispersion of
poly(3,4-ethylene dioxythiophene)-poly(styrene sulfonate) (PEDOT:PSS) was
taken as a high-molecular-weight solution. (2)~A low-molecular-weight
solution was also used that included red and blue light-emitting polymers
dissolved in mesitylene. For the first type of solution, the initial film
thickness was $h_0\approx160\,\mu$m. For the second type of solution, the
film was relatively thin, $h_0\approx1$ to 3\,$\mu$m. The experimental
results showed two process behaviors depending primarily on the film
thickness and molecular weight of the dissolved substance. The dissolved
high-molecular-weight substances, which form a skin on the free surface of
relatively thick films, tend to accumulate in irradiated areas where the
evaporation rate is highest. In contrast, the redistribution of
low-molecular-weight dissolved substances in thin films is caused by the
surface tension gradient resulting from temperature and concentration
differences, which can lead to a decrease in the dissolved substance in the
illuminated areas. In other words, an inverted pattern is obtained.
Figures~\ref{fig:IR_laser_pattern}a and \ref{fig:IR_laser_pattern}c show
the light templates (white color shows illuminated areas). The corresponding
patterns are shown in Figs.~\ref{fig:IR_laser_pattern}b and
\ref{fig:IR_laser_pattern}d. To obtain a text from a red light-emitting
polymer (figure~\ref{fig:IR_laser_pattern}b), the
experimenters~\cite{VieyraSalas2012} applied an inverted light template
(figure~\ref{fig:IR_laser_pattern}a). The thickness of the PEDOT:PSS dried
layer was several microns (figure~\ref{fig:IR_laser_pattern}d). A numerical
simulation of this experiment was performed in~\cite{VieyraSalas2012} (see
Sec.~\ref{sec:models}). Further studies in this field will allow optimizing
the proposed method.

Marangoni flow can also be controlled in real time by a point vapor source.
Changing the location of the vapor source above the droplet in space and
time was proposed in~\cite{Malinowski2018} to achieve the structuring of deposits on the
substrate. The source was a needle with inner diameter of $r_0\approx80$ to
640\,$\mu$m. Dry ethanol was used as vapor. We let $\gamma_1$ denote the
liquid surface tension near the periphery. In the vapor source area, the
surface tension reduces to $\gamma_2$ ($\gamma_1>\gamma_2$). The surface
tension gradient causes the emergence of Marangoni flow that redirects the
transfer of suspended particles. In the absence of ethanol, the capillary
flow forms an annular pattern. In the presence of a vapor source (dry
ethanol), the Marangoni flow arises, and a central spot is formed.

In the experiment in~\cite{Malinowski2018}, 1\,$\mu$L droplets with silica
particles of about 2\,$\mu$m (density of $\rho_\mathrm{Si}=1850$ kg/m$^3$
and the sedimentation rate in water of $v_s\approx2.1$ $\mu$m/s) were used.
The droplet contact angle on glass is $\theta\leq5^\circ$. Measurements
showed that the flow velocity increases as $h$ decreases ($h$ is the
distance from the substrate to needle tip). If $h$ is low, then the flow
velocity $v$ exceeds the sedimentation rate $v_s$ by more than an order of
magnitude. The authors~\cite{Malinowski2018} obtained an analytic
estimate,
$$
v(h)=\frac{h_D}{2\eta R}\left[\gamma(h,R)-\gamma(h,0)\right],
$$
where $\eta$ is the water viscosity, $h_D$ is the droplet height and
$$
\gamma(h,r)=\gamma_W-\frac{\beta}{2}\frac{c_0 r_0}{\sqrt{\left(h-h_D\right)^2+r^2}}.
$$
Here, $c_0$ is the ethanol concentration at the needle tip, and $\beta$ is
the proportionality constant~\cite{Malinowski2018}.

\begin{figure}[ht]
\centering
  \includegraphics[width=0.99\linewidth]{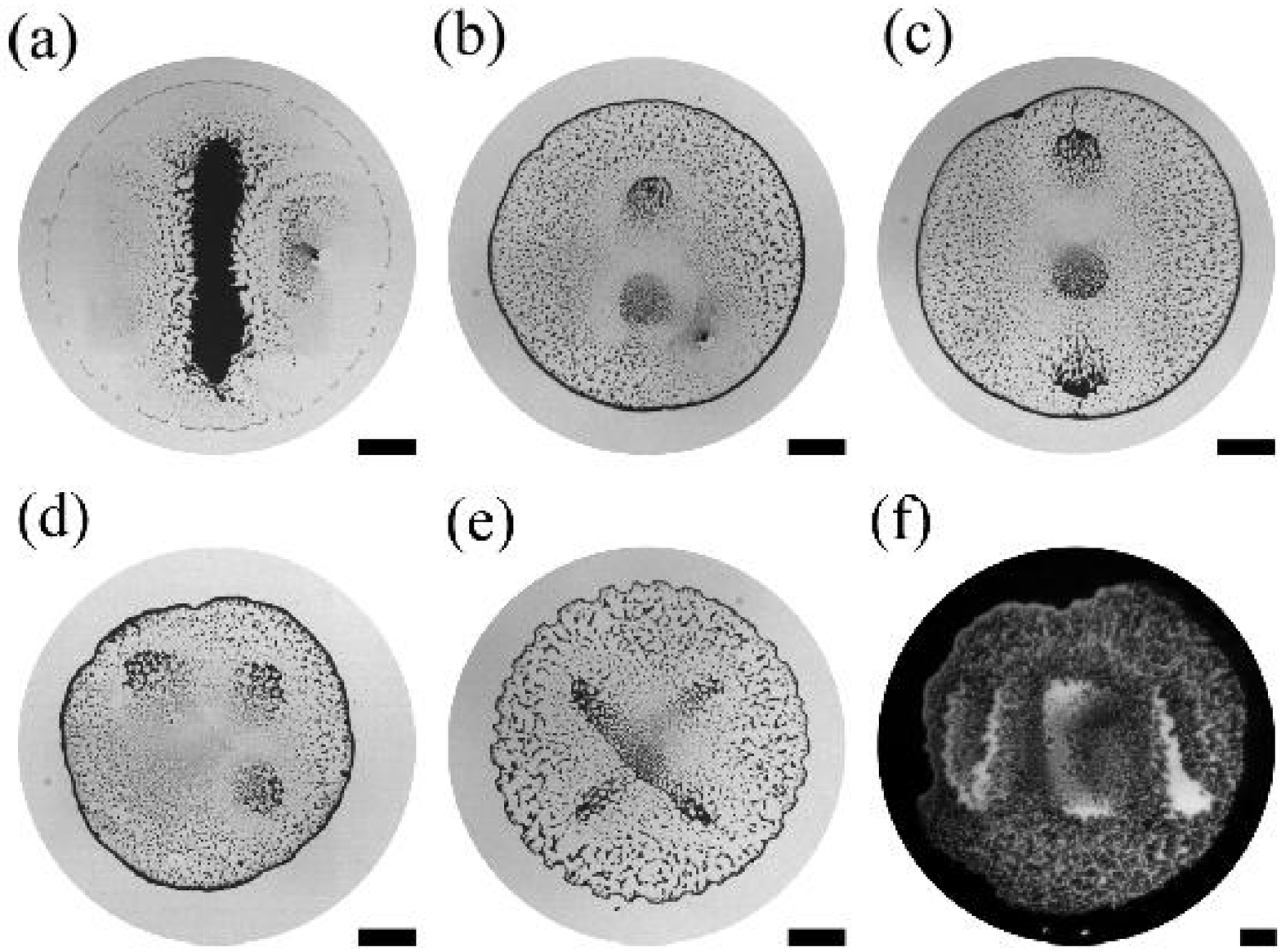}\\
  \caption{Applying a pattern to the surface using dynamic spatial and
  temporal control of the Marangoni flow. Different surface structures of
  deposited colloidal particles after evaporation obtained by shifting an
  ethanol vapor source in (a--c)~one or (d--f)~two spatial directions over
  time: (a)~line, (b)~two dots, (c)~three dots in a line, (d)~three dots in
  a 2D configuration, (e)~cross, and (f)~letters UCL. Scale lines correspond
  to 1~mm. Reprinted with permission from Refs.~\cite{Malinowski2018}.
  Copyright 2018 American Chemical Society.}
  \label{fig:RealTimeControlMarangoniFlows}
\end{figure}

As the vapor source is moved along the droplet surface, the location of the
minimum surface tension can be dynamically shifted. This allows changing the
flow configuration in the mode close to real time.
Figure~\ref{fig:RealTimeControlMarangoniFlows} shows several resulting
patterns corresponding to different needle motion types.

\begin{figure}
\centering
  \includegraphics[width=0.99\linewidth]{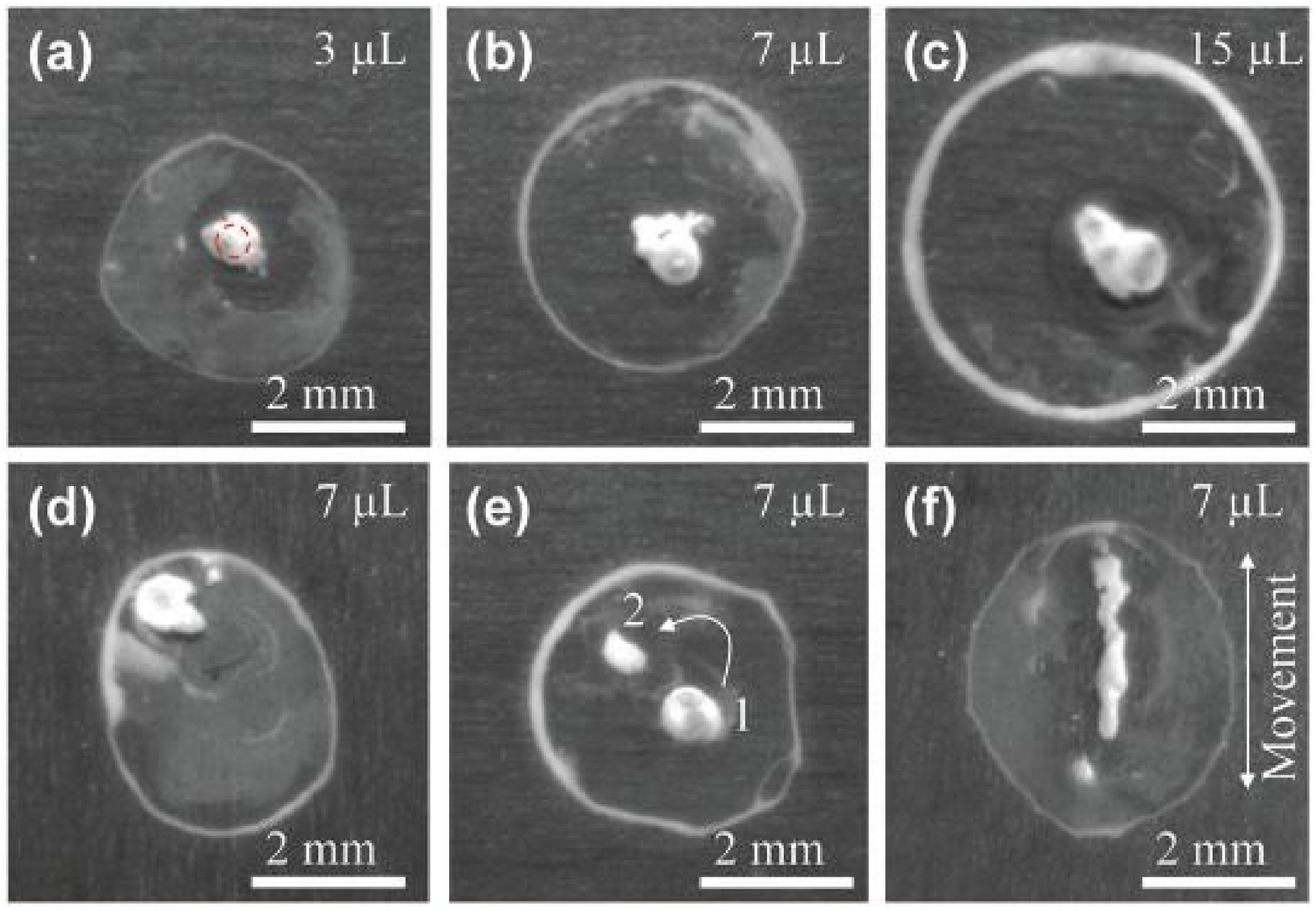}\\
  \caption{Particle deposit formation under exposure to laser beams under
  different conditions: (a--c)~a laser beam is projected to the central part
  of a droplet; (d)~a laser beam is projected near the periphery; (e)~a
  laser beam is projected first to point 1 and after some time to point 2;
  (f)~the substrate was moved relatively to the laser beam with a constant
  speed of 0.12 mm/s. Reproduced from Ref.~\cite{Ta2016} with permission
  from The Royal Society of Chemistry.
}
  \label{fig:LaserOverMovingSubstrate}
\end{figure}

Two years earlier, other researches achieved a similar result by displacing
a substrate with droplet relative to a fixed laser emitter~\cite{Ta2016}
(Fig.~\ref{fig:LaserOverMovingSubstrate}). The locally heated area loses its
liquid due to evaporation relatively faster. Consequently, the compensatory
flow rushes into the illuminated area, capturing and carrying the particles
there. An IR laser with a wavelength of 2.9\,$\mu$m was chosen for the
experiment because water adsorbs the heat of the passing light wave well in
this range. Polystyrene particles of about 0.5\,$\mu$m in size were used.
The authors~\cite{Ta2016} measured the dependence of the evaporation time of
droplets on the radiation power density (53 to 224 W/cm$^2$) for droplets of
2 to 10\,$\mu$L. The size of the formed deposits was also measured as a
function of the beam diameter.  The types of deposits were also defined for
the cases when the laser switches off at some point of time
$t_\mathrm{off}\leq t_\mathrm{fL}$, where $t_\mathrm{fL}$ is the time of the
complete drying of the droplet. The time intervals can be conventionally
divided into three zones: (1)~$t_\mathrm{off}/t_\mathrm{fL}<0.6$,
(2)~$0.6\leq t_\mathrm{off}/t_\mathrm{fL}\leq0.7$, and
(3)~$0.7<t_\mathrm{off}/t_\mathrm{fL}\leq1.0$. Zone~1 corresponds to
coffee-stain or coffee-stain-like structures. Uniform deposits are typical
of zone~2. Inverted coffee stains were observed in zone~3.

The described active methods are preferred in terms of the broad ability to
control the structure of the formed sediment, but they are still inferior in
obtaining high-resolution patterns. These methods currently allow obtaining
structures with a spatial frequency in the submillimeter and millimeter
range. One possible way to solve this problem is to combine different
approaches. Hybrid methods are discussed in detail in the next subsection.

\subsection{Hybrid methods for controlling particle deposition}
\label{subsec:HybridMethods}

Hybrid methods combine flow control through evaporation with other
approaches that are classified as EISA or not. Many variations are possible.
Therefore, we do not try to mention all of them here but give only some
illustrative examples.

Some applications require a uniform and relatively large deposit area of
densely packed particles, for example, for the creation of photonic crystals.
Uniform deposit was obtained in~\cite{Noirjean2017} after directional
evaporation from a Hele-Shaw cell. The colloidal film was in the gap between
two glass plates. All sides except one were covered with varnish. Therefore,
evaporation only occurred in the area of the open cell. The capillary flow
carried the particles and arranged them  into a multilayer structure toward
the opening. The pattern formation was also affected by the capillary
attraction of particles and their electrostatic interaction~\cite{Noirjean2017}.
Weak diffusion was also observed in the system. It is noteworthy that
densely packed and chaotic zones can be distinguished in the structure.
Sometimes, cracks were observed in densely packed zones at the end of the
process. Further study and more detailed knowledge of such processes will
allow understanding ways to improve the properties of the resulting
structure.

\begin{figure}[ht]
\centering
  \includegraphics[width=0.99\linewidth]{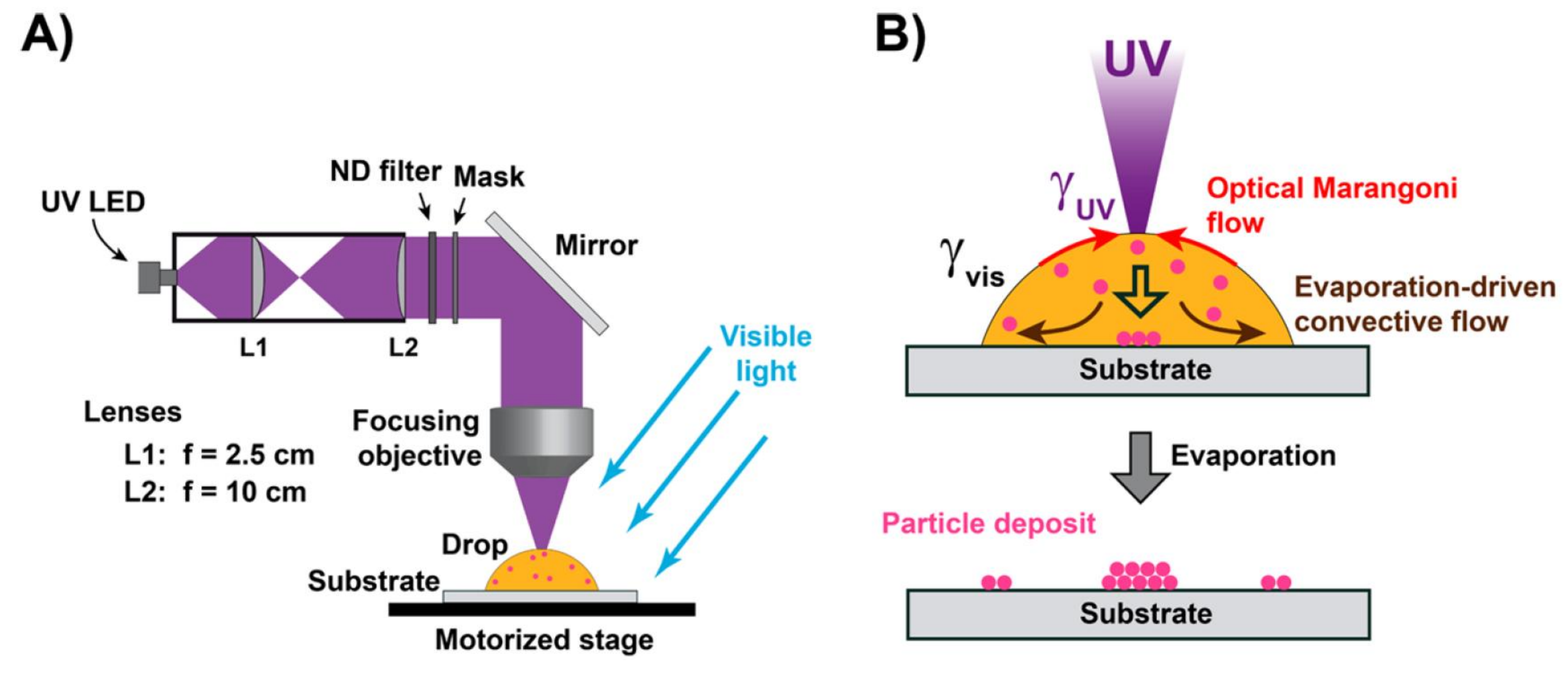}\\
  \caption{(a)~Sketch of the experiment, (b)~flow structure under the UV
  illumination of the central part of the colloidal droplet surface with
  additives of photosensitive surfactants; deposit type after drying.
  Reprinted with permission from Ref.~\cite{Varanakkottu2015}. Copyright
  2015 American Chemical Society.
  }
  \label{fig:eOMA_sketch}
\end{figure}

The method proposed in~\cite{Varanakkottu2015} is based on a combination of
the compensatory flow directed toward the periphery and the optically
induced Marangoni flow directed along the surface from the edge to the
center. The experimental setup is shown schematically in
Fig.~\ref{fig:eOMA_sketch}a. The authors~\cite{Varanakkottu2015} used
photo-sensitive surfactants AzoTab, which increased the water surface
tension under ultraviolet (UV) radiation. As a result, the circulating flow
carried colloidal particles into the central area affected by the UV light
(Fig.~\ref{fig:eOMA_sketch}b). This method was called evaporative optical
Marangoni assembly (eOMA). In Fig.~\ref{fig:eOMA_sketch}b,
$\gamma_{\mathrm{UV}}$ and $\gamma_{\mathrm{vis}}$ are the respective
surface tension factors under exposure to UV and visible light
($\gamma_{\mathrm{UV}}>\gamma_{\mathrm{vis}}$).

\begin{figure}[ht]
\centering
  \includegraphics[width=0.99\linewidth]{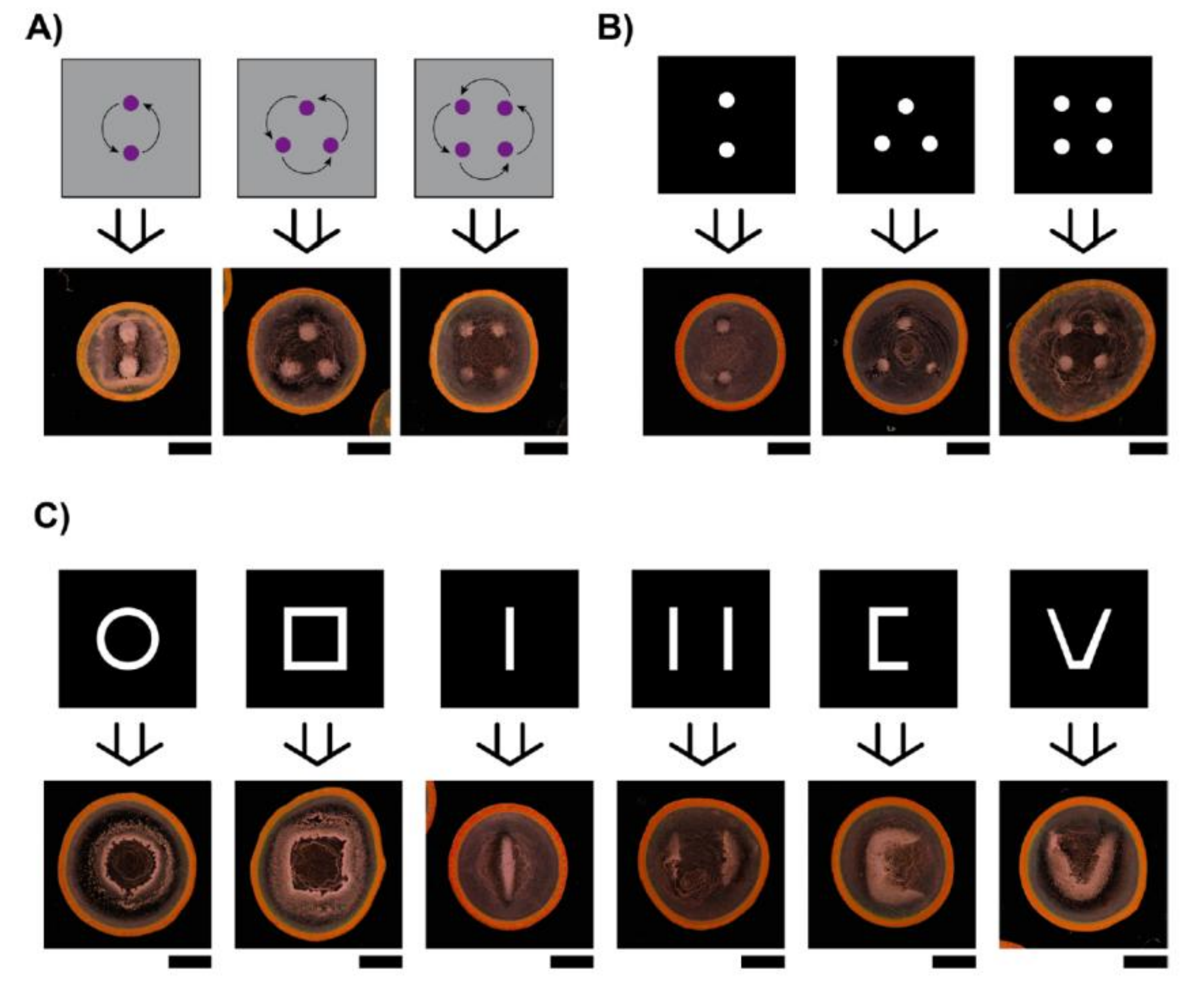}\\
  \caption{Patterns obtained using the eOMA method. (A)~Single-spot scanning,
  (B)~single-spots projected through photomasks, (C)~strips, circles,
  squares, and letters projected through photomasks (scale lines correspond
  to 2~mm). Reprinted with permission from Ref.~\cite{Varanakkottu2015}.
  Copyright 2015 American Chemical Society.
  }
  \label{fig:eOMA_patterns}
\end{figure}

Figure~\ref{fig:eOMA_patterns} shows examples of patterns formed by UV
illumination of some areas of the droplet surfaces during evaporation,
including the use of photo masks with different geometries (points, stripes,
circles, squares, and letters). The droplet volume varied between 12 and
20\,$\mu$L. The authors~\cite{Varanakkottu2015} are confident about the
flexibility of this method. Their experiments proved the applicability of
eOMA to different sizes and charges of particles.

Key parameters of the eOMA method were studied in detail in the subsequent
work~\cite{Anyfantakis2017}. The authors considered the case where a thin
(relative to the droplet base diameter) UV beam was directed to the central
part of the droplet. This allowed obtaining a cylinder-shaped deposit in the
middle of the ring. The following parameters were studied: the concentration
of surfactants and colloidal nanoparticles and the diameter and intensity of
the light beam. These parameters affect the size and volume of light-induced
central particle deposit. In the experiment, 7\,$\mu$L drops were used. The
radius of central deposits varied from 0.1 to 0.7~mm at different values of
the key parameters. Their height varied from 0.5 to 1\,$\mu$m. The volume
varied in the range between 0.05 and 0.3~nL. The Marangoni flow velocity was
assessed based on the formula $U_{\mathrm{Ma}}\approx(h/\eta)\cdot\nabla\gamma$,
where $h$ is the droplet height, $\eta$ is the dynamic viscosity, and
$\nabla\gamma=\gamma_{\mathrm{UV}}-\gamma_{\mathrm{vis}}$. The value of
$\nabla\gamma$ varied in the range of approximately 1 to 12~mN/m depending
on the concentration of surfactants. Measurements showed that the volume of
central deposit increases nonlinearly with increasing colloidal nanoparticle
concentrations. The size of the particles in the experiment was about 30~nm.
The experiment also showed the possibility of forming a cylindrical deposit
in an arbitrary area illuminated by a UV beam. The
authors~\cite{Anyfantakis2017} determined the optimum surfactant
concentration relative to the value of $\nabla\gamma$ varying over time. It
is better to have an initial concentration of $C_s\ll\mathrm{CMC}$, where
$\mathrm{CMC}$ is the critical micellar concentration. Then during most of
the process time, $\nabla\gamma$ takes a relatively high value.

\begin{figure}[ht]
\centering
  \includegraphics[width=0.99\linewidth]{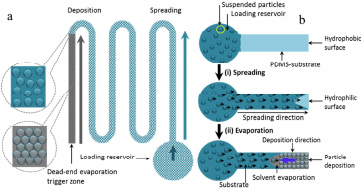}\\
  \caption{(a)~Schematic diagram of the experiment~\cite{Lone2017} with the
  evaporation of liquid from open microchannels. (b)~Three main process
  stages. Reprinted with permission from Ref.~\cite{Lone2017}. Copyright
  2017 American Chemical Society.
  }
  \label{fig:MicrofluidicChannels_sketch}
\end{figure}

Another method combines evaporative lithography and topographic and chemical
surface modification. A method was developed in~\cite{Lone2017} based on
using evaporative lithography in open microfluid channel networks
(Fig.~\ref{fig:MicrofluidicChannels_sketch}a). First, a solid surface was
topographically structured using a photolithography method based on
polydimethylpolysiloxane (PDMS photolithography). For the experiment,
several substrates with different channel geometries were manufactured:
straight, zigzag, and square mesh. Channel widths varied between 5 and
10\,$\mu$m, and the depth was about 5.6\,$\mu$m. The microchannels were
connected to a loading reservoir (a cell in the form of a cavity in the
substrate to locate the droplets). Such a system of cavities in the
substrate was subject to plasma treatment. Hence, the surface wetting
property in these areas changed locally from hydrophobic to hydrophilic.
The multilayer particle structure was obtained in three stages
(Fig.~\ref{fig:MicrofluidicChannels_sketch}b): (1)~Liquid flows quickly
through hydrophilic microchannels after colloidal droplets are placed in the
reservoir. (2)~The flow generated by evaporation carries the particles
toward the dead end of the channel. (3)~The formed deposit boundary slowly
moves from the dead end of the channel toward the reservoir.

Micro- and nanoparticles were used in the experiments~\cite{Lone2017}. The
proposed method will allow for the future production of thin transparent
electric conductive films with the required width of the conductor and the
required gap between adjacent layers. Further development of this approach
can involve its application to polymers, pigments, vesicles, cells,
bacteria, viruses, and other biological molecules to build 2D and 3D
structures. Additional studies are required in this field.

\begin{figure}[ht]
\centering
  \includegraphics[width=0.99\linewidth]{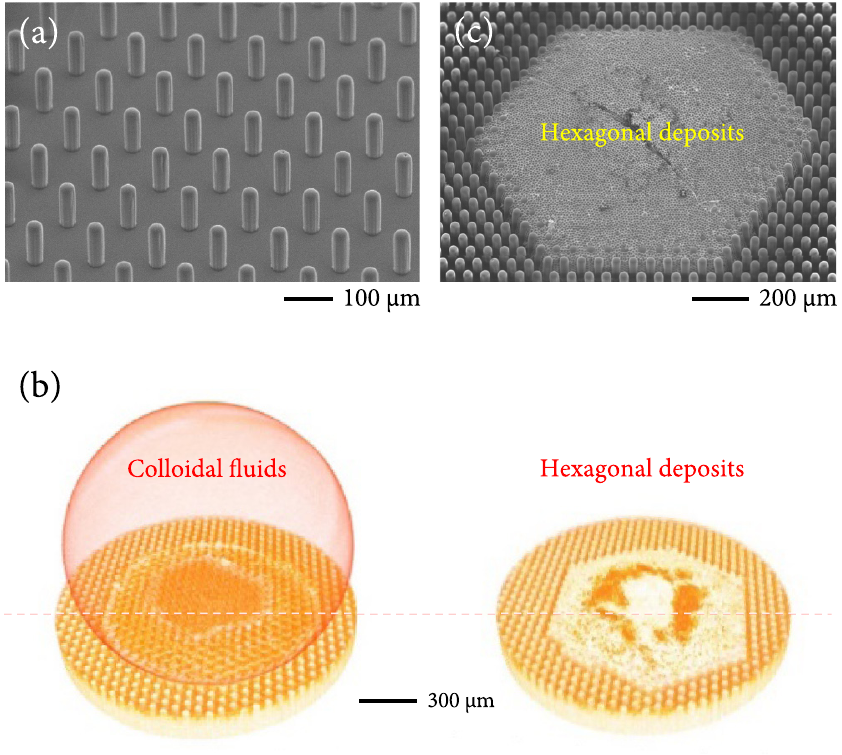}\\
  \caption{Hexagonal deposits: (a)~Image of a hexagonal array of
  micropillars (scanning electron microscopy). (b)~Restored x-ray tomographic
  images of colloidal liquid on a hexagonal array of micropillars (left) and
  hexagonal deposit after evaporation (right). (c)~Image (scanning electron
  microscopy) of the hexagonal deposit on a hexagonal array of micropillars.
  Reproduced with permission from Ref.~\cite{Park2019} (Copyright 2019 by The
  American Physical Society).
  }
  \label{fig:HexagonalDepositsOnMicropillarArrays}
\end{figure}

Another experiment showed the possibility of forming polygonal deposits when
a colloidal droplet dries on a microstructured substrate~\cite{Park2019}.
Soft lithography followed by etching was used to fabricate such a substrate
with an array of micropillars (Fig.~\ref{fig:HexagonalDepositsOnMicropillarArrays}a).
The droplet volume in the experiment~\cite{Park2019} was about 2\,$\mu$L.
Polystyrene particles of different diameters were used (0.1, 0.5, 1.0, 10,
20\,$\mu$m). Structures with a different pitch between the pillars (30, 90,
120, 140\,$\mu$m) were used with fixed values of the pillar diameter
(30\,$\mu$m) and height (100\,$\mu$m). We note that the structured deposits
obtained using this method contained no cracks
(Fig.~\ref{fig:HexagonalDepositsOnMicropillarArrays}b,c), which is important
for inkjet printing, for example. The phenomenon of hexagonal pattern
formation requires clarification, and this issue remains
open~\cite{Park2019}.

A combination of colloidal bead self-assembly and evaporative lithography
was proposed in~\cite{Vakarelski2009}. A liquid with gold nanoparticles
(approximately 20~nm in diameter) was added to a monolayer of densely packed
polystyrene microspheres (50 or 100\,$\mu$m in size) on the substrate. As a
result, a network of liquid bridges formed in the cavities between the
microspheres. Packing of particles can be controlled by adding a surfactant
to the solution~\cite{Vakarelski200913311}. The mechanism of stabilizing the
liquid bridges was studied in~\cite{Vakarelski2013}. After the microspheres
are removed with adhesive tape, the network remains on the substrate. During
slow evaporation, nanoparticles do not drift to the contact line because
diffusion prevails over convection. After evaporative assembly, microwire
patterns are formed from nanoparticles. The width of the individual wires
ranges approximately between 2 and 4\,$\mu$m. Their height is about 1 to
2\,$\mu$m. The authors called this method ``bridging
lithography''~\cite{Vakarelski2009}. The resulting network structure can be
used as a transparent conductive coating. Measurements showed transparency
exceeding 80\% and resistance per square from 0.5 to 1.0
$\Omega/\Box$~\cite{Vakarelski2009}. A mask template placed on the substrate
made using the photolithography method can be used instead of
microspheres~\cite{Tang2010}. To obtain quantum dots in the form of
nanorings, a combination of colloidal lithography and capillary lithography
can be used~\cite{Chen2009}. This method was called ``evaporative
templating.'' In the drying film, microspheres are formed as a monolayer on
the substrate. Nanoparticles occupy the pores between these spheres.
In the process of evaporation, nanoparticles are attracted to the
microspheres by capillary action. The latter are in turn removed from the
surface. As a result, many individual nanorings with a diameter varying from
a few nanometers to several hundred nanometers remain on the
substrate~\cite{Chen2009}. The authors of~\cite{Wedershoven201892} proposed
to combine die-coating technology, an EISA method based on a capillary
bridge (stick-slip motion of contact line), and active evaporation control
due to local air flow.

Hybrid methods can be both active and passive
(Fig.~\ref{fig:EISA_classification}). Development in this direction can be
associated with combining other effects with evaporative lithography. In
particular, the transfer of particles can be affected by an electrowetting
effect~\cite{Eral2011}, magnetic~\cite{Josten2017,Darras2019,Saroj2019,Al-Milaji2019}
and acoustic~\cite{Rudenko2010} waves, gravity~\cite{Li2018} and
centrifugal force (spin coating)~\cite{Hoggan2004,Huang2010}, porous
substrate imbibition~\cite{Mampallil2018,Al-Milaji2019}, salt
additives~\cite{Saroj2018,Sun2019,Homede2020}, etc. The emergence of new
methods and the wide interest in evaporative lithography are due to a large
number of relevant applications in various fields of human activity.

\subsection{Evaporative lithography applications}
\label{applications}

Such promising applications as the formation of transparent and conductive
structures for the purpose of optical electronics and
microelectronics~\cite{Lone2017,Vakarelski2009,Zhang2013,Layani2014},
mass transfer in nanoliter cells applicable to labs-on-a-chip~\cite{Rieger2003},
and polygonal deposits without cracks for inkjet printing~\cite{Park2019}
have already been mentioned above. We now consider the prospects of applying
evaporative lithography in various areas in detail.

One of the promising directions in micro- and optoelectronics is the use of
organic materials. For example, organic transistors are created by drying
droplets with crystals after application on a dielectric substrate by an
inkjet printing method~\cite{Lim2009}. Active control over deposition
in an evaporating solution using infrared radiation can be used
in an organic light-emitting diode (OLED) creation process~\cite{VieyraSalas2012}.
Sediment formation in a drying droplet is one possible approach for
producing P-OLED (polymer-organic light-emitting diode)~\cite{Eales2015} and
QLED (quantum dot light-emitting diode) displays~\cite{Chen2009,ParkY2019,Jiang2016}.
Another promising application in optical electronics is the creation of
flexible thin-film photovoltaic elements~\cite{Deng2015}. A modification of
EISA with a tilted movable substrate was used in~\cite{Deng2015} to obtain
cyclic thin films of perovskite nanowires.

A transparent conductive structure was obtained in~\cite{Layani2009}, where
an inkjet printing process was used to form an array of droplets on a
substrate. After the droplets dried, the substrate was left with an array of
ring deposits. The procedure was repeated several times. Water ink with a
silver nanoparticle content of about 0.5\% of the solution weight was used.
The particle diameter was about 5--20~nm. The resulting patterns of
overlapping rings are characterized by good transparency (95\%) and
resistance per square (specific resistance of 0.5~cm$^2$ is approximately
4~$\Omega/\Box$). The ring had a width of about 10\,$\mu$m, a height of
about 300~nm, and a diameter of about 150\,$\mu$m.

This technology was later applied to carbon nanotubes, which are a good
alternative to traditional materials (metal oxides) for optoelectronic
devices~\cite{Shimoni2014}. In~\cite{Layani2011}, liquid droplets were
applied to a metal mesh placed over a plastic substrate. After drying, the
metal mesh was removed. A mesh-shaped deposit of silver nanoparticles
remained on the substrate. After the chemical sintering of the particles by
HCl vapor treatment at room temperature, a transparent electrically
conductive flexible film was obtained~\cite{Layani2011}. Silver nanowire
structures were also obtained in~\cite{Wang2017}.

\begin{figure}[ht]
\centering
  \includegraphics[width=0.95\linewidth]{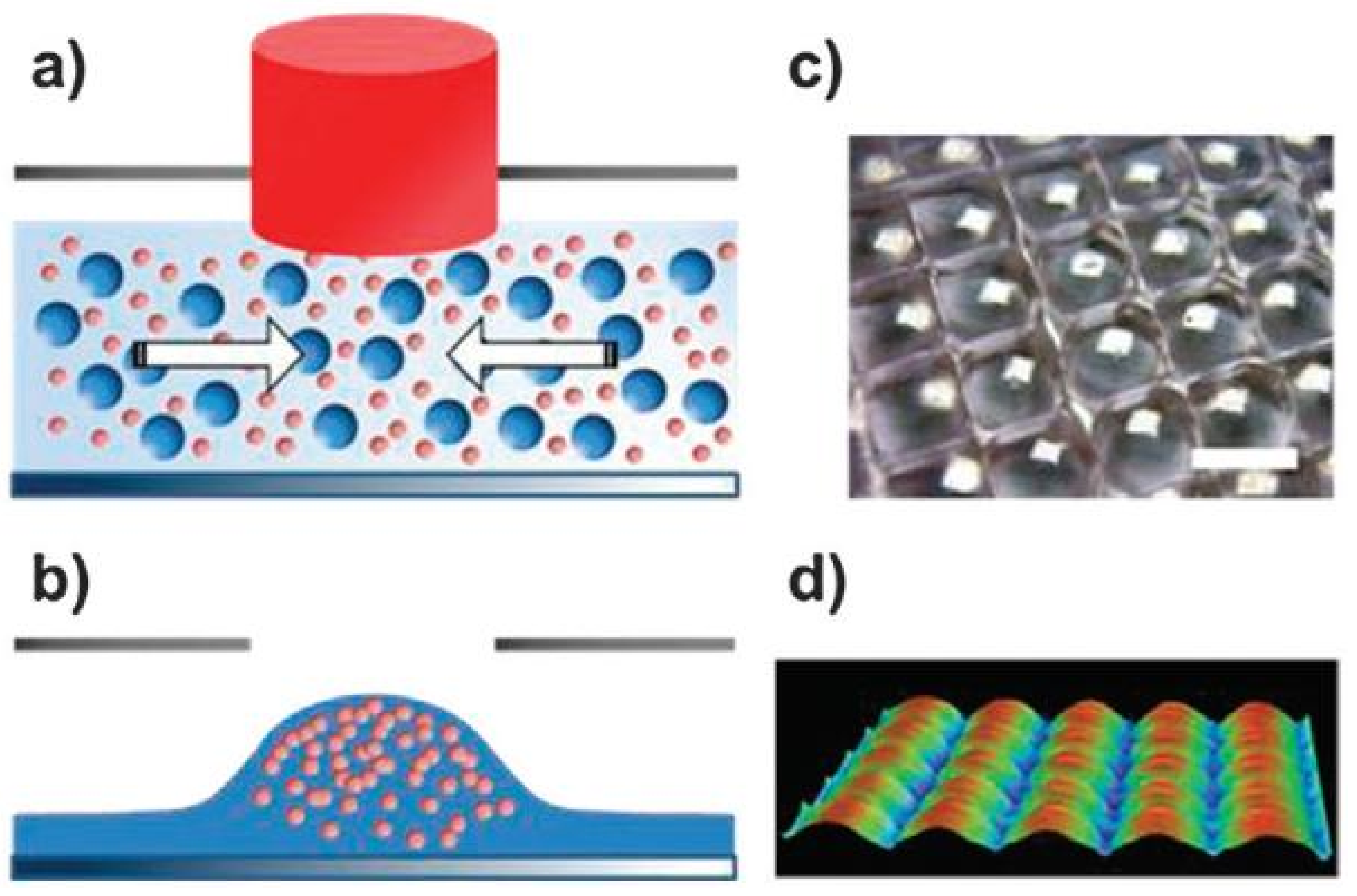}\\
  \caption{Schematic illustration of the IRAEL process: (a)~Solution of
  polymeric acrylic particles (blue) and gold nanoparticles (red).
  (b)~Polymeric particles sinter and form a solid coating containing gold
  nanoparticles. (c)~Photograph of a solid film with periodic properties
  (scale 2~mm). (d)~Three-dimensional surface profilometry. Reproduced from
  Ref.~\cite{Utgenannt2013} with permission from The Royal Society of
  Chemistry.
  }
 \label{fig:GoldNanospheresInPolymerCoating}
\end{figure}

Plasmonic metamaterials can be manufactured using the IRAEL
method~\cite{Utgenannt2013,Utgenannt2016}. A mask and an IR lamp were used
in~\cite{Utgenannt2013}. After the polymer solution with gold nanoparticles
(10~nm in diameter) dried, a textured glass layer was formed
(Fig.~\ref{fig:GoldNanospheresInPolymerCoating}). In the resulting solid
film, nanoparticles were concentrated in convex areas of the polymer that
formed under the holes in the mask. Different sizes of polymer particles
(160, 280, 400~nm) were used in the experiment~\cite{Utgenannt2013}. The
proposed approach to organizing inorganic nanoparticles in a polymeric
coating is one-step and suitable for regulating optical properties on a
macroscale with a period of several millimeters. By reducing the size of
the polymer particles, this technology can also be used to obtain structures
whose optical properties change with a period of about 10~microns.

In~\cite{Utgenannt2016}, a 2D plasmonic structure of gold nanoparticles was
obtained on the surface of a 3D photonic crystal without using a mask. The
authors~\cite{Utgenannt2016} constructed a model based on the Langevin
equation to describe Brownian particle dynamics. The LAMMPS library was used
in the calculations. Polymeric particles on the liquid free surface play the
role of a template. Rapid evaporation under light heating results in an
accumulation of particles near the two-phase liquid--air boundary. This is
the case when the Peclet number $\mathrm{Pe_{evap}}>1$ ($\mathrm{Pe_{evap}}=
HE/D_s$, $E$ is the free boundary movement velocity during evaporation,
$D_s$ is the particle diffusion coefficient, and $H$ is the initial film
thickness). The process takes about seven minutes. As a result of diffusion,
gold nanoparticles are arranged in voids between polymer particles (338~nm
in diameter) near a free surface. This approach allows for obtaining
periodic nano- and microscale patterns in a short time.

Superlattices of structured iron oxide nanospheres coated with oleic acid
can also be obtained by evaporating droplets on a silicon wafer~\cite{Josten2017}.
In experiments~\cite{Josten2017,Yang2018}, the evaporation rate was
controlled by placing the wafer with a droplet inside a special cell.
Moreover, particles were additionally exposed to a weak magnetic
field~\cite{Josten2017}. In~\cite{Yang2018}, a staircase superstructure of
nanocuboids (golden core coated with silver enclosure) was obtained in a
ring-shaped deposit.

\begin{figure}[htb]
\centering
  \includegraphics[width=0.99\linewidth]{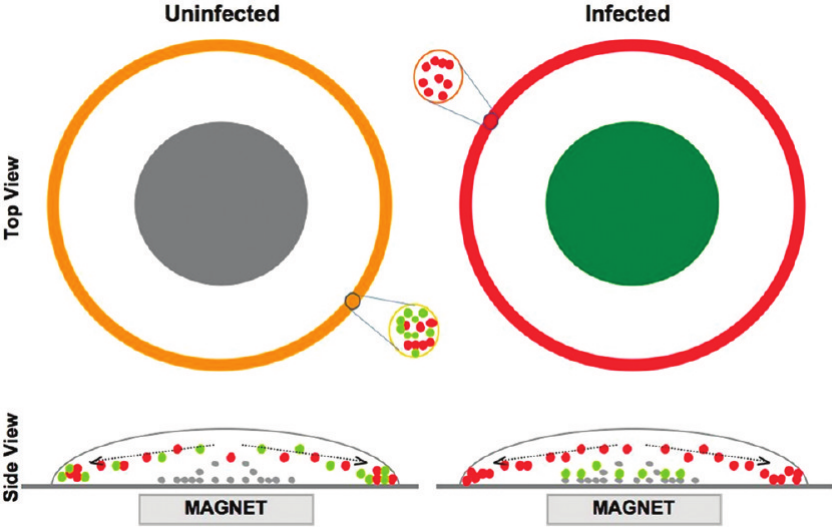}\\
  \caption{Coffee ring assay schematic. Reprinted with permission from
  Ref.~\cite{Trantum2011}. Copyright 2011 American Chemical Society.
}
  \label{fig:CoffeeRingAssay}
\end{figure}

The formation of the deposit pattern can be used to diagnose diseases in
health care. For example, a simple, low-cost method based on the disturbance
of the coffee-ring pattern by a biomarker was described in~\cite{Trantum2011},
where this diagnostic method was tested in an example application of
diagnosing malaria. In the experiment, the biological sample was mixed with
a solution containing three kinds of particles. Of these, the surface of two
kinds of particles is able to interact with the biomarker of the disease.
These are magnetic particles (grey) and biomarker indicator particles
(green). The third type is inert particles (red). Biomarkers in the solution
bind the surfaces of functionalized particles (magnetic and indicators).
Gray particles are deposited in the central part due to the magnetic field
(Fig.~\ref{fig:CoffeeRingAssay}). The flow of liquid carries the red
particles to the periphery of the drying droplet. In the absence of a
biomarker, the green particles are also carried to the forming ring deposit.
The combination of red and green particles gives the ring a yellow color,
indicating a negative test result. If the result is positive, the ring color
is red. In addition, the central spot inside the ring becomes green as the
indicators are deposited together with magnetic particles. Such structures
are easy to interpret.

Another method based on the coffee ring effect allows detecting protein
molecules (e.g., thrombin) in a sample~\cite{Wen2013}. One more application
in health care is the formation of microneedles on a plate for local safe
and painless intracutaneous injections of drugs~\cite{Mansoor2011,Mansoor2012}.
We also mention biosensors used in environment and food safety
analysis~\cite{Zhou2020}.

\begin{figure}[htb]
\centering
  \includegraphics[width=0.99\linewidth]{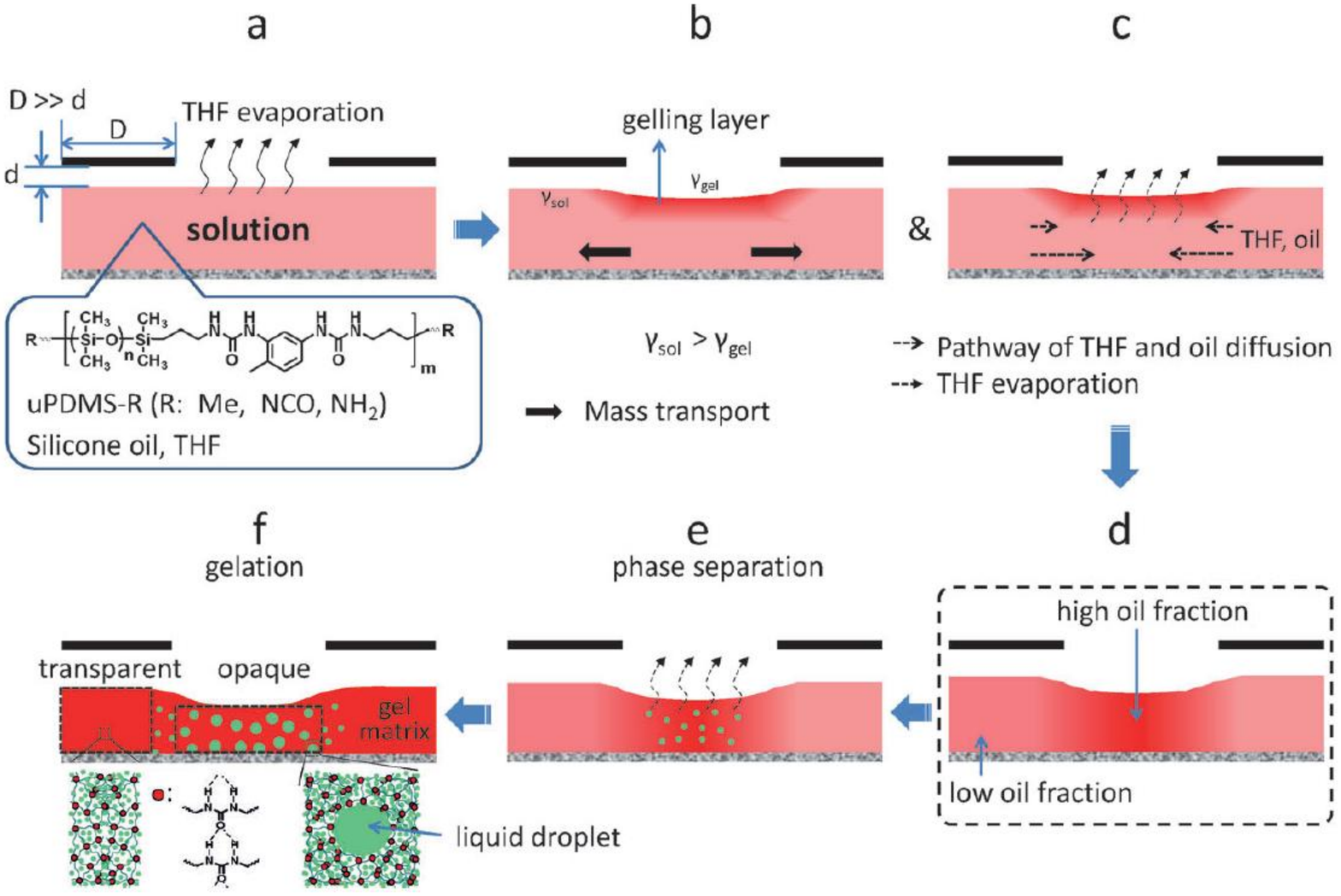}\\
  \caption{Control of localization of liquid droplets in a gel matrix using
  evaporative lithography: (a)~A homogeneous solution consisting of
  copolymer, silicone oil, and volatile tetrahydrofuran. The solution film
  is applied to a flat substrate and covered with a mask on top. (b)~The
  volatile liquid evaporates rapidly in open areas resulting in a surface
  tension gradient and hence mass transfer from open to closed areas.
  (c)~Tetrahydrofuran diffuses from closed to open areas as does oil.
  (d)~Nonuniform distribution of silicone oil. (e)~Phase separation due to
  excess oil in open areas resulting in formation of silicone droplets.
  (f)~Gelation and formation of a structured coating. Reproduced with
  permission from Ref.~\cite{Zhao2016}, Copyright 2016 WILEY-VCH Verlag
  GmbH \& Co. KGaA.}
  \label{fig:oil_droplets_in_gel_matrix}
\end{figure}

Localized fluid inclusions provide solid materials with multiple functions
such as self-recovery, secretion, and controlled mechanical properties in a
spatially controlled mode. An experiment~\cite{Zhao2016} suggested a method
for selectively localizing liquid drops in a supramolecular gel obtained
from the solution using evaporative lithography
(Fig.~\ref{fig:oil_droplets_in_gel_matrix}). The formation of areas
saturated with suspended droplets occurs in the free evaporation area where
the nonvolatile liquid is concentrated. The phases are thus separated before
gel formation. A homogeneous gel matrix forms in the areas with hindered
evaporation. The subareas underneath the mask openings and in the closed
areas were characterized by different functional properties: slipperiness,
self-healing, transparency, etc.~\cite{Zhao2016}. There is no mathematical
description of the process in~\cite{Zhao2016}, and the schematic legend of
Fig.~\ref{fig:oil_droplets_in_gel_matrix} therefore raises additional
questions. For example, how does the compensatory flow transfer relate to
the redistribution of the substance by the solutal Marangoni flow? Analytic
estimates and mathematical modeling will allow further understanding of the
mechanisms of this process in more detail. We discuss the current state of
development of mathematical models of processes related to evaporative
lithography in Sec.~\ref{sec:models}.

\subsection{Dimensionless parameters in evaporative lithography}

When studying heat and mass transfer processes, dimensionless
parameters controlling the nature of these processes are of natural
interest (see e.g.~\cite{bookHMT}). For physical processes occurring in evaporating
sessile droplets, there are additional dimensionless characteristics whose
effect is very significant. The many parameters whose roles were analyzed
in this respect in~\cite{Larson20141538} include the Bond number, the
buoyancy number, the ratio of substrate to fluid thermal conductivities,
and many others. We present the main dimensionless parameters additionally
arising in evaporative lithography problems in Table~\ref{tab:dimensionalGroups}.

\begin{table*}[htb]
  \centering
  \caption{Dimensionless parameters in evaporative lithography}%
    \begin{tabular}{|c|p{20em}|c|}
    \hline
    Dimensionless parameter & Definition & Typical values \\
    \hline
    $h_g/P$ & Ratio of gap $h_g$ to pitch $P$ & 0--1.25 \\
    \hline
    $\phi$ & Initial volume fraction of particles & 0--0.64 \\
    \hline
    $P/d_h$ & Ratio of pitch $P$ to hole diameter $d_h$ & 2--10 \\
    \hline
    $\Delta J/J_\mathrm{free}$ & Ratio of vapor flux density difference $\Delta J$
    to freely evaporating film vapor flux density $J_\mathrm{free}$ & 0--10 \\
    \hline
    $\mathrm{Pe_{evap}}=HE/D_s$ & Peclet number related to the evaporation-induced interface motion & 10$^{-2}$--10$^8$ \\
    \hline
    $\psi=V_M/V_E$ & Ratio of Marangoni flow velocity $V_M$ to compensatory flow velocity $V_E$ & 0.01--2000 \\
    \hline
    $a_\mu/a_\mathrm{nano}$ & Ratio of microparticle radius $a_\mu$ to nanoparticle radius $a_\mathrm{nano}$ & 2--1000 \\
    \hline
    $R/H$ & Ratio of rod radius $R$ to rod--film gap $H$ & 0.5--4 \\
    \hline
    $L_i/D_c$ & Ratio of interdroplet distance $L_i$ to contact diameter $D_c$ & 1--$\infty$ \\
    \hline
    $D_\mathrm{beam}/D_c$ & Ratio of beam diameter $D_\mathrm{beam}$ to droplet size $D_c$ & 0.12--0.22 \\
    \hline
    $t_\mathrm{off}/t_\mathrm{fL}$ & Ratio of irradiation time $t_\mathrm{off}$ to evaporation time $t_\mathrm{fL}$ & 0--1 \\
    \hline
    $\mathrm{Re_{air}}=\rho_\mathrm{air}u_{\max}R_1/\mu_\mathrm{air}$ & Jet Reynolds number ($\rho_\mathrm{air}$ and $\mu_\mathrm{air}$ are the air density and viscosity) & 0--10 \\
    \hline
    $\mathrm{We}=\rho_\mathrm{air}u_{\max}^2R_1/\gamma$ & Weber number ($\gamma$ is the liquid surface tension) & 0--1 \\
    \hline
    $\mathrm{Pe}=V_{\max}R_1/D_s$ & Peclet number related to the lateral fluid flow ($V_{\max}$ is the lateral flow velocity) & 10$^{-2}$--10$^8$ \\
    \hline
    $r_0/h$ & Ratio of inner needle radius $r_0$ to needle--substrate distance $h$ & 0.04--0.32 \\
    \hline
    $C_s/\mathrm{CMC}$ & Ratio of initial surfactant concentration $C_s$ to critical micellar concentration $\mathrm{CMC}$ & 0--1 \\
    \hline
    \end{tabular}%
  \label{tab:dimensionalGroups}%
\end{table*}

The complex structure of boundaries arising in such problems produces a
large number of geometric dimensionless parameters that affect the
experimental outcome. As already discussed in Sec.~\ref{sub:PassiveMethods},
an increasing $h_g/P$ ratio leads to a decrease in
$\Delta h$~\cite{Harris2007,Routh2011}. Weak particle segregation is
observed for $h_g/P\geq0.3$ (the value of $\Delta h$ is small)~\cite{Harris2007}.
The values of $\phi$ determine the type of sediment (discrete elements in
the masked sediment for $\phi<0.07$ or continuous textured sediment layer
for $\phi>0.07$)~\cite{Harris2007}. The value of $\Delta h$ depends on
$\phi$ nonlinearly. It begins to decrease for $\phi>0.35$~\cite{Routh2011}.
At a rather high $\phi>0.22$, Marangoni flow can be
suppressed~\cite{HarrisLewis2008}. The phase transition of the liquid film
to the solid phase occurs for $\phi\geq0.64$~\cite{Harris2007}. An increase
of $P/d_h$ causes an increase in distance between adjacent elements of
sediments~\cite{Harris2007}. The diameter of the elements of nanoparticle
sediments inside pores formed by microparticles corresponds to the size of a
hole in the mask at $a_\mu/a_\mathrm{nano}\geq7$~\cite{Harris2009}. An
increase of $R/H$ leads to an increase in the evaporation rate gradient
above the film in the rod area~\cite{Parneix2010}. An adjacent droplet has
no effect when the values of $L_i/D_c$ are rather high. Conversely, the
maximum effect of one droplet on the mass transfer process in another is
observed as $L_i/D_c\to1$~\cite{Hegde2018}. An increase in
$D_\mathrm{beam}/D_c$ results in an increase in sediment size relative to
the drop size~\cite{Ta2016,Anyfantakis2017}. Reducing the value of $r_0/h$
reduces the spot sediment area until it completely disappears at
$r_0/h=0.04$~\cite{Malinowski2018}. At a relatively high value
($r_0/h\approx0.32$), the spot is strongly smeared over the contact area of the
droplet with the substrate. Particle transfer to the irradiated area
is more efficient for $C_s/\mathrm{CMC}\ll1$~\cite{Anyfantakis2017}. In this
case, the surface tension difference $\nabla\gamma=\nabla\gamma(C_s)$ takes
relatively high values for most of the drop evaporation time, which leads to
a strong Marangoni flow.

Parameters such as $\mathrm{Pe}$, $\mathrm{Pe_{evap}}$, and $\psi=V_M/V_E$
characterize heat and mass transfer processes in the liquid phase. For
$\mathrm{Pe}\gg1$, flow transport dominates, which results in obtaining a
structured sediment. For $\mathrm{Pe}<1$, diffusion transport dominates,
which results in uniform sedimentation~\cite{Wedershoven2018}. Skin
formation occurs for $\mathrm{Pe_{evap}}\gg1$~\cite{Utgenannt2016,Wedershoven2018,Okuzono2006}.
Marangoni flow prevails for $\psi>1$~\cite{Routh2011}. In the case of small
initial particle concentrations, this can lead to an inverted sediment
structure~\cite{HarrisLewis2008}. If compensatory flow prevails ($\psi<1$),
then the particles are transported into areas under holes in the mask.

The parameter $\Delta J/J_\mathrm{free}$ characterizes mass transfer in the
air phase and affects the hydrodynamics. As it increases, $\Delta h$ also
increases~\cite{Harris2007}. Here, $J_\mathrm{free}$ is the vapor flux
density above the film with no mask, and the difference between the vapor
flux densities in an area closed with a mask and an open area (under a
hole) is $\Delta J=J_{\max}-J_{\min}$. During exposure by air flow to
evaporating film at Reynolds number values $\mathrm{Re_{air}}\gg1$, a dip
may be observed in the film until a rupture
occurs~\cite{Wedershoven2018,Berendsen2012}. In such cases, particle
transfer  is no longer controlled by adjusting the vapor concentration above
the free surface of the liquid. There are often several dry spots if the
value of the Weber number is approximately unity  or greater~\cite{Berendsen2012}.
These parameters are restricted in evaporative lithography
($\mathrm{Re_{air}}$ near unity and $\mathrm{We}\ll1$).

Parameters such as $t_\mathrm{off}/t_\mathrm{fL}$ can also be relevant in an
experiment. This parameter determines the portion of the process time
controlled by external radiation. The rest of the time, the process occurs
naturally (without external influences). Hence, the value of
$t_\mathrm{off}/t_\mathrm{fL}$ affects final sediment shape~\cite{Ta2016}.
An annular sediment is observed for $t_\mathrm{off}/t_\mathrm{fL}<0.6$. A
value in the range from 0.6 to 0.7 results in a uniform sediment. Inverted
deposition is obtained for $t_\mathrm{off}/t_\mathrm{fL}>0.7$.

The listed parameters facilitate a comparative analysis of experiments for
which they are important. For example, the mass transfer process depends on
the main key parameter $\Delta J/J_\mathrm{free}$ in experiments with a
mask~\cite{Harris2007} and with a composite substrate~\cite{Cavadini201324}
(Sec.~\ref{sub:PassiveMethods}). In the case of a composite substrate,
$J_\mathrm{free}$ corresponds to the vapor flux density in the area where
the thermal properties of the substrate were not changed previously (Teflon
coating, etc.). The parameter $\Delta J/J_\mathrm{free}$ can be expressed
theoretically in terms of the ratio of thermal conductivities
$(\lambda_2-\lambda_1)/\lambda_1$ (Fig.~\ref{fig:surfaceCurvatureReasons})
for the experiment~\cite{Cavadini201324}.

\section{Mathematical modeling of heat and mass transfer in droplets and films under nonuniform evaporation}
\label{sec:models}

\subsection{Preliminary remarks}

Although evaporative lithography has by now evolved as an independent field of
research, the development of relevant applied methods and theoretical
approaches has often been inextricably linked with the development of other
studies of the evaporation of liquid layers and droplets. Some of the
relevant studies are or may have been predecessors to this field, while
others may be useful for its further development. We present a few examples.

The model in~\cite{Popov2005} describes the geometry of depositions during
their formation in evaporating colloidal droplets on a flat substrate.  Some
assumptions of the model, such as independence of evaporation from free
droplet surface from the availability of dissolved substances, were
discussed by the author in detail. The transport of colloidal particles to
the contact line in the case of an asymmetric evaporating colloidal droplet
in which the contact line contains an acute angle was studied
in~\cite{Zheng2005} in the approximation of homogeneous evaporation. The
formation of a structured sediment upon drying of a colloidal droplet was
modeled in~\cite{Kolegov2019} with the effect of the capillary attraction of
particles taken into account. Based on the simulations, the physical
mechanisms of forming individual chains of particles inside an annular
sediment were analyzed.

To describe the spatial distribution of colloidal particles and hydrodynamic
flows in an evaporating sessile droplet and also the structure of the
sediment after drying, the obtained experimental results were analyzed
in~\cite{Bhardwaj2009} using numerical simulation based on the finite
element method of the joint solution of the Navier--Stokes equations in a
liquid, the heat conduction equation in the droplet and substrate, the
equation of vapor diffusion in air, and the convection--diffusion equation
to account for the transfer of particles in the droplet. A negligible
influence of colloidal particles on fluid hydrodynamics was assumed, and an
axisymmetric solution was considered (2D model). During evaporation, the
solid phase area can propagate when the critical concentration of the
solution achieves dense packing of the particles. The contact line becomes
depinned when the wetting angle decreases below a certain value.
In~\cite{Bhardwaj2010}, electrostatic and van der Waals interactions between
colloidal particles and the solid substrate were added to the model using DLVO
theory. This leads to DLVO forces in the convection--diffusion equation. The
DLVO interaction forces depend on the pH value of the solution, for which
several different values were used. Depending on the parameter values,
single-ring deposits, a homogeneous spot, or accumulation of particles in
the center of the dried-up droplet were obtained as a result of the
calculations~\cite{Bhardwaj2009,Bhardwaj2010}.

Periodic pinning and depinning of a moving contact line and formation of
structured sediments on glass plates during evaporation in a vertical
Hele-Shaw cell were studied experimentally and theoretically
in~\cite{Bodiguel2009,Bodiguel2010}. The obtained expression for the pinning
force demonstrates that it is determined by the geometry of the growing
sediments and the force of gravity.

A similar problem statement in which the meniscus of a thin film of a
colloidal fluid is in contact with a moving substrate was considered
in~\cite{Doumenc2010,Doumenc2013}. For the computer simulation, the
solution of the hydrodynamics and the dissolved substance transport were
calculated in the framework of the lubrication approximation, and
evaporation was also taken into account using the diffusion model. The
influence of capillary forces on the meniscus shape was not accounted
for in~\cite{Doumenc2010,Doumenc2013}, because the film thickness remained
rather large. Solutal Marangoni flow was considered in~\cite{Doumenc2013};
the dependence of the saturated vapor density on the concentration of the
dissolved substance in solution was taken into account in \cite{Doumenc2010}.
These works numerically investigated the modes of deposition depending on
the speed of the substrate and studied periodic deposit structures for
several parameter values~\cite{Doumenc2010,Doumenc2013}. Practically the
same problem statement was studied in~\cite{Frastia2011,Frastia2012}, which
also used the hydrodynamic model, but the role of the wetting and periodic
pinning/depinning effects of the moving contact line in the formation of
various lines from deposits was also investigated. The authors analyzed the
dependence of sediment properties on numerous experiment parameters.

A simplified model of the evaporation of a droplet of a polymer solution
with the compensatory flow, diffusion of the dissolved substance, and the
separation of liquid and gel parts of the droplet taken into account was
proposed in~\cite{Okuzono2009}. It used the lubrication approximation,
while the droplet surface was found quasistatically and was approximated by
a parabolic curve with time-dependent factors. The vapor flux density was
assumed to be constant in the area of the solution and equal to zero in the
area of the gel part. The model allows distinguishing between final deposits
of the basin, crater, and mound types depending on the problem parameters.
The influence of the surface gel film on the evaporation rate of the solvent
and on the collective diffusion coefficient was studied in~\cite{Okuzono2008}
for polymer solutions drying on a solid substrate.

There are other models that could be applied to problems of evaporative
lithography with some modifications~\cite{Zang2019}. We next consider
approaches to modeling processes of evaporation lithography developed to
date.

\subsection{Modeling of evaporative lithography processes}

Several models have been developed for evaporative lithography. These models
are based on different approaches, for example, the kinematic approach or
the lubrication approximation. Different methods are used to solve these
problems, for example, the finite-difference method (FDM) or the finite
element method (FEM). As a rule, processes are considered in liquids
(single-phase models without film solidification). Two-phase models are used
less frequently because they are complex. In reality, different phase
transitions occur as the volatile substance is lost, for example, a liquid--vapor
phase transition, glass formation during melting and cooling of a polymer
sediment, a sol--gel transition, and salt crystal formation as the solution
concentration increases. Some models take multiphase systems into account.
Certain flow types dominate in different complex systems, for example,
Marangoni or Rayleigh--B\'enard convection, capillary flow, etc. Additional
effects including sedimentation, coagulation, etc., must often be considered.

We consider the most used approaches to mathematical modeling of different
processes in evaporative lithography and the corresponding approximations.
We classify the methods that do not use the fluid motion equations obtained
from the momentum conservation law under the kinematic approach. In such
studies, the flow velocity is expressed based on the mass conservation law,
while the motion and shape of the two-phase liquid--air interface are
defined based on various assumptions using an approximation. The term
``kinematic approach'' was selected because such models do not consider the
causes of the flow (forces, pressure gradients, etc.), nor do they
explicitly include such parameters as fluid viscosity, for instance. The
equilibrium free surface of thin droplets is often approximated at each
instant using the parabolic approximation where the parameters of the
parabola depend on the time and the evaporation rate. A flat free surface is
sometimes used for films, with the film height linearly decreasing over time
proportionally to the evaporation rate. Local free surface curvatures and
the deviation from the equilibrium shape are possible in real systems. They
can be caused both by internal and external forces. Such curvatures can
often be almost imperceptible, but it is precisely these curvatures that
cause capillary flows in microdroplets or flows caused by the hydrostatic
pressure gradient in macrodroplets. The kinematic models do not explain this
phenomenon.

Models can be classified into continuous and semidiscrete~\cite{Lebedev-Stepanov2013}.
In semidiscrete models, the fluid flow is described continuously using the
mass and momentum conservation laws, while the dynamics of each individual
particle is considered. In the continuous approach, particle redistribution
is described based on the convection--diffusion
equation~\cite{Okuzono2009,Bhardwaj2009} in terms of concentration. The
discrete approach that explicitly describes individual particles and liquid
molecules is computationally intensive and hence rarely
used~\cite{Lebedev-Stepanov2013}. In the semidiscrete approach, the particle
dynamics are often simulated using dissipative particle
dynamics~\cite{LebedevStepanov2013} or the Monte Carlo
method~\cite{Andac2019,Kolegov2019}. The continuous approach is based on the
Navier--Stokes equations
\begin{equation}\label{eq:motionEquation}
  \rho \left(
    \frac{\partial \mathbf{v}}{\partial t} +
    \left(\mathbf{v}\cdot \nabla\right)\mathbf{v}
  \right) =
  - \nabla P + \mu\, \Delta \mathbf{v}
\end{equation}
and the continuity equation
\begin{equation}\label{eq:continuityEquation}
  \nabla\cdot \mathbf{v} = 0.
\end{equation}
Here, $\rho$ is the liquid density, $\mu$ is the viscosity, $P$ is the
pressure, and $\mathbf{v}=(\mathbf u,w)$ is the flow velocity vector, where
$\mathbf u$ is the velocity vector in a plane section parallel to the
substrate and $w$ is the velocity component perpendicular to the substrate.
We note that gravity is disregarded in~\eqref{eq:motionEquation}. This
applies when the typical size of the liquid is less than the capillary
length. In an axisymmetric case, it is often convenient to write
Eqs.~\eqref{eq:motionEquation} and~\eqref{eq:continuityEquation} in terms of
the stream function and vorticity, thus eliminating the unknown function
$P$~\cite{Barash2009}.

If the fluid layer is relatively thin, then using the lubrication
approximation is often acceptable~\cite{Oron1997,Fischer2002}. System of
equations~\eqref{eq:motionEquation}, \eqref{eq:continuityEquation} is then
simplified and becomes
\begin{multline}\label{eq:LubricationApproximationNavierStokes}
\mu \frac{\partial^2 {\mathbf u}}{\partial z^2} - \tilde\nabla P =
0,\quad -\frac{\partial P}{\partial z} = 0,\quad \tilde\nabla
{\mathbf u} + \frac{\partial w}{\partial z} =0,
\end{multline}
where $\tilde\nabla$ is the horizontal projection of the gradient. The
formulation of the boundary conditions varies in different studies and
depends on the particular problem. If two expressions with pressure
from~\eqref{eq:LubricationApproximationNavierStokes} are integrated with the
particular boundary conditions taken into account, then explicit equations
for $\mathbf u$ and $w$ can be obtained~\cite{Fischer2002}. The flow
velocity averaged over the fluid layer thickness is
$$
\bar{\mathbf u} = \frac{1}{h}\int \limits_0^h \mathbf u\, dz.
$$
The value of $z=h$ corresponds to the position of the liquid--air interface.

It is often appropriate to use the lubrication approximation together with
the kinematic approach. Analytic expressions for the fluid flow velocity
that are applicable for a sufficiently small film height or sufficiently
small droplet contact angle can then be obtained. In~\cite{Hu2005,HuLarson2005},
analytic velocity expressions were obtained for a thin droplet, and the
velocity field was compared with the results of numerical simulation. The
compensatory flow was studied in~\cite{Hu2005}, and Marangoni flow was
studied in~\cite{HuLarson2005}. If the droplet surface is approximated by a
spherical cap instead of a parabola, then analytic expressions can be
obtained for the lubrication approximation, which are applicable with good
accuracy for a significantly broader range of contact angles, while the
expression complexity barely increases~\cite{Barash2016}. If the kinematic
approach is used, even more accurate but cumbersome analytic approximations
for the velocity field can be obtained without using the lubrication
approximation at all~\cite{Barash2016}. We next discuss works related to the
mathematical modeling of processes in evaporative lithography.

The model in~\cite{RouthRussel1998} describes the mass transfer of a
nonuniformly dried aqueous latex film. A lubrication approximation was used
in~\cite{RouthRussel1998} to derive the basic equations of the model. The
evaporation rate was assumed to be constant in the open areas, and no
evaporation was assumed in the masked areas. Compensatory flows appear as a
result of deformation of the free surface and evaporation. In the vertical
direction, the Peclet number is equal to zero, and it is hence believed that
the particles are evenly distributed throughout the thickness of the thin
film. In the horizontal direction, the Peclet number is equal to infinity,
and particle diffusion is hence not taken into account. Using the mass
conservation law, the authors obtained the equation for the time-dependent
film thickness
\begin{equation}\label{eq:ConservationLawForFilmThickness}
    \frac{\partial h}{\partial t} + \tilde\nabla \left(h\bar{\mathbf u}\right) = - \frac{J l}{\rho},
\end{equation}
where the vapor flux density $\tilde J$ is equal to~0 or $J_0$ depending on
the spatial coordinate ($J_0>0$ is a constant). It mimics the effect of the
mask on evaporation. The notation $l=\sqrt{1+(\tilde\nabla h)^2}$ is used
in~(\ref{eq:ConservationLawForFilmThickness}). In~\cite{RouthRussel1998},
$l=1$ for the flat free surface of the liquid and also the flow rate
$\bar{u}=\sigma/(3\mu)\,h^2\,\partial^3 h/\partial x^3$ obtained
from~\eqref{eq:LubricationApproximationNavierStokes} were used in the case
of the single horizontal direction $x$ by considering a boundary condition
for $P$ at the surface using the capillary pressure ($\sigma$ is surface
tension). The particle transfer equation was formulated in terms of the mass
fraction $\phi$,
\begin{equation}\label{eq:TransferEquation}
\frac{\partial (h \phi)}{\partial t} + \tilde\nabla\left( \phi h
\bar{\mathbf u} \right) = 0.
\end{equation}
Along with the capillary flow in the liquid film, the fluid filtration in
the solid film, where the mass fraction of particles reaches a critical
value $\phi_m =0.64$ corresponding to the random dense packing of spheres,
was taken into account in~\cite{RouthRussel1998}. Darcy's law was only used
to derive the motion of the liquid front of the film, which moves as the
solid part dries completely. An additional equation that allows explicitly
defining the film front with dense packing of particles was also obtained
in~\cite{RouthRussel1998}. The problem was solved numerically using
the fourth-order Runge--Kutta method. The authors reported a natural experiment
in~\cite{RouthRussel1998} (Fig.~\ref{fig:RouthRussel_exper98}). Its results
agreed qualitatively with the results of their numerical solution. In the
final film profile, there were dips in the area where the evaporation was
blocked by the mask. This is due to an outflow of mass from such an area to
the areas where evaporation occurs. The capillary flow compensates for the
loss of vaporized mass.

As previously mentioned (see Sec.~\ref{sub:PassiveMethods}), Deegan et
al.~\cite{Deegan2000} considered three modes of evaporation in their
experiments. They applied their mathematical model to only one of these
modes, where the evaporation rate increases towards the three-phase
boundary. The given explanation of the coffee-ring effect was  based on a
simple kinematic model~\cite{Deegan2000}. In this model, the dissolved
substance is carried to the contact line by compensatory fluid flow. The
droplet profile $h$ was approximated by a spherical cap shape. This droplet
geometry is typical of small liquid volumes where capillary forces
dominate the gravitational forces. Experimental measurements~\cite{Deegan2000}
confirm that the approximation of $h$ by the equilibrium form is allowable.
The height-averaged radial velocity of the compensatory flow caused by
evaporation in~\cite{Deegan2000} was expressed from the mass conservation
law~\eqref{eq:ConservationLawForFilmThickness},
$$
\bar u(r,t) = - \frac{1}{\rho r h} \int\limits_0^r dr\,r \left( J(r, t)
\sqrt{1+ \left( \frac{\partial h}{\partial r} \right)^2} + \rho
\frac{\partial h}{\partial t}  \right).
$$
Particle transfer equation~\eqref{eq:TransferEquation} written in
cylindrical coordinates was also included in the model~\cite{Deegan2000}.
The diffusion-limited evaporation model was used to calculate the vapor flux
density $J$. If evaporation can be regarded as an adiabatic process in the
sense that the vapor concentration adjusts sufficiently rapidly to a change
in the drop size and shape, then the Laplace equation with the corresponding
boundary conditions can be used as an approximation to the diffusion
equation for vapor in air. An analytic approximation based on the exact
solution of the Laplace equation obtained in toroidal coordinates was
proposed. The problem was shown to be mathematically equivalent to the
known problem of the electrostatic potential of a charged conductor with a
shape defined by two intersecting spheres. The disadvantage of the
approximation $J\approx J_0f(\lambda)\left[1-(r/R)^2\right]^{-\lambda}$
proposed in~\cite{Deegan2000} is that $J\to\infty$ at the three-phase
boundary. Here. $R$ is the contact line radius, and $\lambda$ and $f$ can
depend on $\theta$~\cite{Deegan2000}. An alternative mechanism for the
coffee-ring effect was proposed in~\cite{Kang2016}, where the authors showed
numerically that the fluid flow in an evaporating droplet can entrain the
particles to the liquid--gas interface, where the particles can be captured
by the surface. The particles are further carried on along the interface
until they reach the contact line.

Another model based on the lubrication approximation was used for a thin
axisymmetric droplet of diluted colloidal solution~\cite{Fischer2002}.
As in~\cite{RouthRussel1998,Deegan2000}, thermal processes were not taken
into account. Mass transfer in liquids was considered at the initial stage
long before the gel phase formation (single-phase model). The droplet shape
was determined by an equation derived from the mass conservation law for the
solution~\eqref{eq:ConservationLawForFilmThickness}. For a thin drop, a
simple case of $l\approx1$ was considered. The dynamics of mass fraction of
particles in space and time were described by the transfer equation. The
flow velocity field and radial velocity averaged over the liquid layer
thickness were derived from~\eqref{eq:LubricationApproximationNavierStokes}
taking into account capillary pressure in the absence of Marangoni flow. The
purpose was to describe the process mathematically for three different
evaporation modes corresponding to the experiments~\cite{Deegan2000}
(Fig.~\ref{fig:DeeganExperiment}). The case where the vapor flux density
dominates near the periphery was described by the model law
$$
J=\frac{J_0}{K+h(r,t)/h_0}\left(1-\mathrm{e}^{-A\,(r-R)^2/R^2}\right),
$$
where the parameter $A$ sets the decrease rate of $J$ near the three-phase
boundary and the nonequilibrium parameter $K$ determines the difference
between the evaporation rate in the droplet center and near the periphery
($K\to0$ in the case of a rapidly evaporating liquid and $K\to\infty$ for a
nonvolatile liquid). Here, $J_0\approx k\,\Delta T/(L\,h_0)$ is the vapor
flux density at the droplet apex, where $k$ is the thermal conductivity of
the liquid, $L$ is the latent heat of vaporization, $h_0=h(0,0)$ is the
initial droplet height, and $\Delta T$ is the difference between the
saturation temperature and the substrate temperature. The case where
evaporation is uniform was described by the model law
$$
J=\frac{J_0}{4h(0,t)/h_0}\left(1-\tanh\left[A\,(r-r_0)/R\right]\right),
$$
where the parameter $r_0$ defines the point near the three-phase boundary
where $J$ tends to zero. The third case, where evaporation prevails
at the symmetry axis, was described by the model law
$$
J=\frac{2J_0}{h(0,t)/h_0}\mathrm{e}^{-A r^2/R^2}.
$$
The three model evaporation laws in~\cite{Fischer2002} are approximations
derived from different considerations describing the process at a
qualitative level. These approximations take into account that the pinning
occurs due to the accumulation of particles near the three-phase
boundary~\cite{weon2013}. Therefore, in these model laws, $J\to0$ as
$r\to R$. In reality, the vapor flux density strongly depends on the
particle concentration in the solution~\cite{Okuzono2010,Tarasevich2011,Kim2018}.
This is especially true for concentrated solutions. Further, the numerically
implicit Gear method was applied to the problem at different values of the
main parameters (evaporation number and capillary number). The velocity
field, the dynamics of the particle distribution, the evolution of the
droplet surface shape, the change in liquid volume, and the contact angle
dynamics during evaporation were obtained in~\cite{Fischer2002}.

The effect of external laser beam heating on the transport of nanoparticles
and liquid in a sessile droplet on a flat horizontal substrate was studied
numerically in~\cite{Dietzel2005}. Parameters such as the laser radiation
intensity, beam diameter, solution absorption factor, and thermal
conductivity of the substrate were studied. The model takes the
thermocapillary flow and droplet shape change under the influence of laser
radiation into account. In the absence of radiation, the model demonstrated
the droplet shape as a spherical segment, which is typical of a size that is
smaller than the capillary length. When exposed to the laser, the shape of
the droplet was curved, and a dimple appeared in the center. An increase in
temperature in local areas leads to a reduction in surface tension. When the
diameter of the beam is comparable to the size of the droplet base, the
surface shape becomes unstable, and ripples appear in the form of waves. The
model in~\cite{Dietzel2005} is based on the Navier--Stokes equations written
in Lagrangian coordinates. In addition, the possible coagulation of
particles was also taken into account. The Galerkin finite element method
was used in~\cite{Dietzel2005} to solve the problem. Unfortunately, this
model does not describe the final structure of the deposit after the
exposure to laser radiation and the complete evaporation of the droplet.

\begin{figure}[ht]
\centering
  \includegraphics[width=0.95\linewidth]{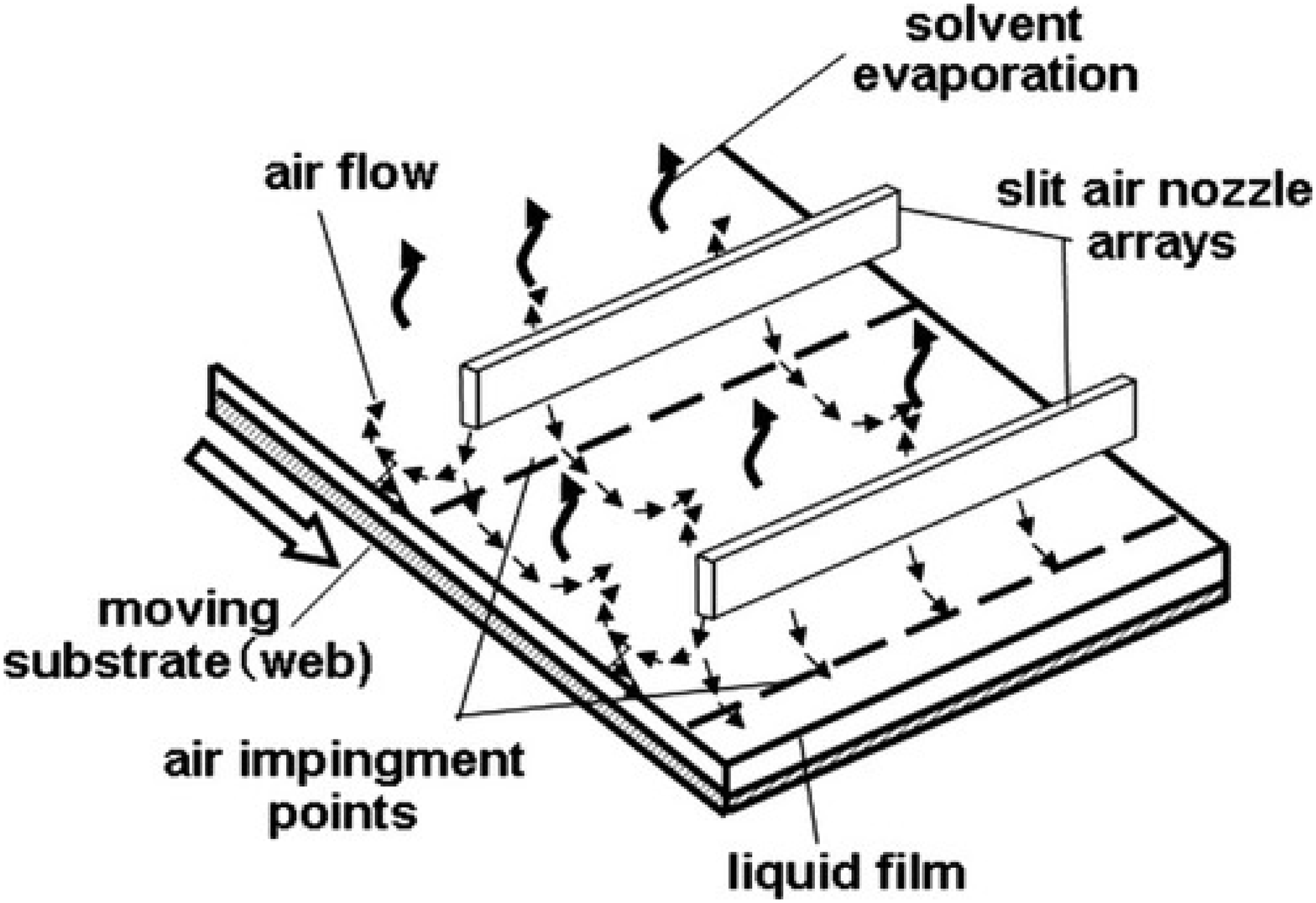}\\
  \caption{Coated film on a moving substrate is periodically blown by air
  from slit nozzles. Reprinted with permission from Ref.~\cite{Yamamura2009}.
  Copyright 2009 American Institute of Chemical Engineers (AIChE).
}
 \label{fig:movingSubstrateAirFlow}
\end{figure}

A method based on the effect of periodic air flow on the evaporation of a
polymer solution film was proposed in~\cite{Yamamura2009}. A liquid with
polymer particles is applied to a substrate above which there is an array of
airflow sources (Fig.~\ref{fig:movingSubstrateAirFlow}). The system has
several geometric parameters: the nozzle hole width $B$, the distance $2W$
between adjacent nozzles, and the distance $H$ between the initial position of
the two-phase boundary and a nozzle. The substrate moves at a certain speed
$U$, which is also one of the system parameters. This method was studied
numerically~\cite{Yamamura2009}. The mathematical model is based on the
lubrication approximation, and the expression for velocity takes not only
the capillary flow but also the Marangoni flow into account,
$$
\bar{u}=-\frac{h^2}{3\mu}\frac{\partial P}{\partial x}+\frac{h}{2\mu}\frac{d\sigma}{dx}.
$$
Here, the Laplace pressure is $P=-\sigma\,\partial^2h/\partial x^2$. In the
model~\cite{Yamamura2009}, the viscosity $\mu$ and surface tension $\sigma$
are expressed through empirical dependences on the temperature and particle
concentration. The vapor flow density is described by a semiempirical
dependence that takes the difference in vapor pressure near and away from
the free film surface into account. The substrate motion is simulated by
changing the Biot number and Nusselt number over time. In~\cite{Yamamura2009},
an implicit finite-difference scheme on an irregular mesh was used to solve
this problem. The discretized scheme was solved by the Thomas algorithm.
Numerical calculations showed that a moving coating (polymer solution) is
subject to asymmetric changes in thickness, which increases and levels
sequentially.

The effect of external laser heating on the formation of the relief coating
was simulated in~\cite{VieyraSalas2012}. A two-dimensional area was
considered, which can be divided into several subareas: the substrate, the
liquid heatsink under the substrate, the film on the substrate, and the air
area above the film. The problem was solved in several stages. (1)~The
temperature distribution due to IR illumination was calculated. (2)~The
solvent vapor transport was calculated, and the vapor flux density was
determined. (3)~The distribution of the dissolved substance in the liquid
layer was simulated. The first stage involved solving the equation
$$
\rho(x,y)\,c(x,y)\,\mathbf v\cdot\nabla T=\nabla\cdot[k(x,y)\,\nabla T+q_v],
$$
where $T$ is the temperature, $k$ is the liquid heat conductivity, and $c$
is the specific heat capacity. The cooling flow velocity $\mathbf v=[u(y),0]$
in the heatsink was determined explicitly ($\mathbf v=0$ in the substrate
and the film). The volume heat flux density is $q_v=a\,I(y)\,H(x)$, where
$a$ is the absorption coefficient, $I$ is the radiation intensity, and $H$
is the Heaviside function for modeling of selective heating of a certain
area. In the second stage, after a stationary temperature profile on the
film surface is found, the convective and diffusion transfer of solvent
vapor in the air area above the film was simulated. This stage is described
by three equations: the equation of motion for the airflow velocity
$\mathbf{v}$ and the convection--diffusion equations for air temperature $T$
and for the vapor concentration $s$. The vapor flux density at the two-phase
boundary is defined as $J(x)=D_v\,\mathbf{n}\cdot\nabla s$, where
$\mathbf{n}$ is the normal vector to the free surface and $D_v$ is the vapor
diffusion coefficient. In the third stage, the mass transfer of the
substance in the liquid was calculated. Different approaches were used for
relatively thin and thick liquid layers. In the first case, the model
includes a one-dimensional particle mass fraction convection--diffusion
equation averaged over the liquid layer thickness and the free surface
dynamics equation based on the mass conservation law. The height-averaged
radial velocity is found using the lubrication approximation. The
hydrostatic and capillary pressure and also the dependence of the surface
tension on the concentration and temperature are taken into account. In the
second case, the model includes a two-dimensional convection--diffusion
equation and the equation of motion. The arbitrary Lagrangian--Eulerian
(ALE) scheme was used in~\cite{VieyraSalas2012} to describe the deformation
of the liquid--air interface. The results of the numerical calculations
in~\cite{VieyraSalas2012} showed the dynamics of film thickness and
particle concentration in space and time. The dependence of the final
structure on the width of the laser beam was also shown.

\begin{figure}[ht]
\centering
  \includegraphics[width=0.95\linewidth]{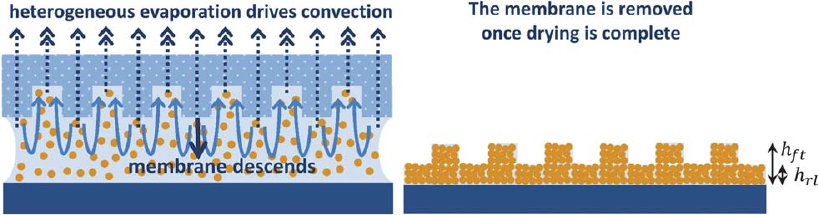}\\
  \caption{Schematic representation of a film evaporation method under a
  topographically patterned membrane. Reproduced from Ref.~\cite{Arshad2013}
  with permission from The Royal Society of Chemistry.
}
 \label{fig:PatternedMembrane}
\end{figure}

The evaporation of a colloidal film under a solvent-permeable membrane with
periodic thicknesses was studied theoretically in~\cite{Arshad2013}. The
essence of the proposed method is that evaporation occurs more rapidly in
those subregions where the thickness of the membrane is thinner
(Fig.~\ref{fig:PatternedMembrane}).

Mass transfer was numerically calculated in the proposed
system~\cite{Arshad2013}. The model includes convection--diffusion and
Navier--Stokes equations. An incompressible liquid is assumed, and the
continuity equation is hence stationary. The driving factor in the system is
the osmotic pressure gradient, which is part of the equation of motion. The
vapor flux density is defined using the formula $J=\bar p\Delta P/d$, where
$\bar p$ is the membrane permeability factor, $\Delta P$ is the partial
pressure driving force, and $d$ is the membrane thickness. The calculation
results showed that the concentration of particles in the area where the
thickness of the membrane is smaller increases. As a result, a solid film of
periodic thickness is obtained. The key parameters for regulating the final
pattern are the geometric characteristics of the membrane along with the
initial film thickness and the solution concentration.

The evaporation of an aqueous latex film under a mask and an IR lamp (see
Sec.~\ref{sec:experiments})~\cite{Routh2011} was simulated
in~\cite{Kolegov2018113}. The model includes a convection--diffusion
equation with a source term,
\begin{equation}\label{eq:convection_diffusion}
\frac{\partial C}{\partial t}+\bar u\frac{\partial C}{\partial x}=
\frac{D_s}{h} \frac{\partial}{\partial x}\left(h\frac{\partial
C}{\partial x}\right) +\frac{J C l}{\rho h},
\end{equation}
mass conservation law~\eqref{eq:ConservationLawForFilmThickness}, the
heat transfer equation for the liquid
\begin{multline}\label{eq:HeatTransferInLiquid}
    \frac{\partial T}{\partial  t}  + \bar u
    \frac{\partial T}{\partial x} =\frac{1}{c \rho h} \frac{\partial
    }{\partial x}\left(k h \frac{\partial T}{\partial x} \right)
    -\frac{J l}{\rho h} ( \frac{L}{c}- T)+\\ + \alpha \frac{ T_s -
    T}{c \rho h} + q\frac{ 1- \exp (-a h)
    }{c \rho h} H(X_h - x),
\end{multline}
and the heat conduction equation for the substrate
\begin{multline}\label{eq:ThermalConductivityInSubstrate}
    \frac{\partial T_s}{\partial t} = \frac{1}{c_s \rho_s}
    \frac{\partial }{\partial x}\left(k_s \frac{\partial T_s}{\partial
    x} \right) + \alpha \frac{T - T_s}{c_s \rho_s h_s} +\\
    +q \frac{\exp(-a h)}{c_s \rho_s h_s} H(X_h - x).
\end{multline}
Here, $D_s$ is the diffusion coefficient, $2X_h$ is the mask hole width,
$\alpha$ is the convective heat transfer coefficient, $L$ is the latent heat
of evaporation, and $q$ is the surface density of the radiation power. The
particle mass fraction $C$ and the temperature $T$ are height-averaged
functions. In~\eqref{eq:ThermalConductivityInSubstrate}, the subscript $s$
denotes the substrate. The local impact of the emitter is described using
the Heaviside function $H$. The model~\cite{Kolegov2018113} accounts for
such effects as diffusive and convective particle transport, thermodiffusion
of heat in the liquid and substrate, convective heat transfer in the liquid,
film cooling by evaporation, liquid and substrate heating by IR light, and
heat transfer between the liquid and substrate.
Equations~\eqref{eq:convection_diffusion}, \eqref{eq:HeatTransferInLiquid},
and \eqref{eq:ThermalConductivityInSubstrate} are derived from the
conservation laws when considering mass and heat balance in an elementary
volume. The capillary flow rate $\bar{u}=\sigma/(3\mu)\,h^2\,\partial^3 h/\partial x^3$
is determined with the lubrication approximation in~\cite{Kolegov2018113}
($l=1$ in this case). An empirical formula is used to describe the
dependence of the viscosity $\mu$ on the particle concentration.
In~\cite{Kolegov2018113}, the liquid--solid phase transition is assumed to
occur when a critical solution concentration $C_g=0.7$ is reached. A
theoretical formula taking the dependence of the evaporation rate on the
liquid temperature into account is used for the vapor flux density. This
formula was modified to also take the effects of particle concentration and
the mask hole size into account. As a result of numerical calculations, the
dynamics of the temperature, the particle mass fraction, the liquid--glass
boundary, the vapor flux density, and the free surface of the film were
obtained~\cite{Kolegov2018113}. The final thickness of the solid patterned
film agrees well with the experiment~\cite{Routh2011}. At the end of the
process, the value of $h$ varies between 3 and 80\,$\mu$m. The capillary
flow velocity initially increases because the vapor flux density increases.
Subsequently, it begins to decrease as the viscosity increases.

\begin{figure}[ht]
\centering
  \includegraphics[width=0.95\linewidth]{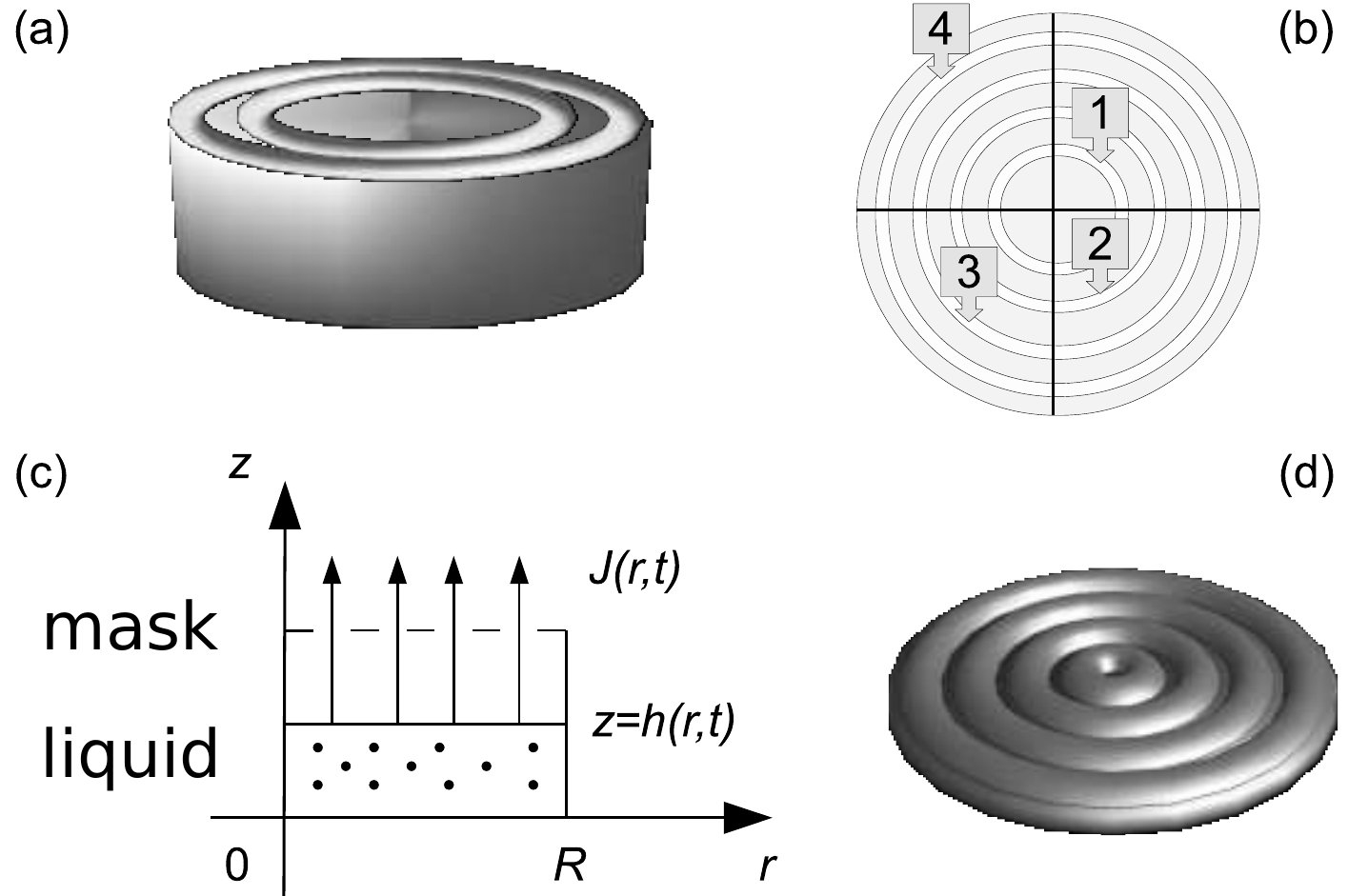}\\
\caption{(a)~Cylindrical cell covered with a mask on top. (b)~A mask with
four annular holes. (c)~Problem domain. (d)~The expected result for particle
redistribution and formation of a solid phase in the form of concentric
rings~\cite{Kolegov201424}.
}
 \label{fig:MaskWithConcentricHoles}
\end{figure}

A method for obtaining concentric deposits of colloidal particles was
proposed in~\cite{Kolegov201424} (Fig.~\ref{fig:MaskWithConcentricHoles}).
A mask geometry with annular holes was used. The method was studied
theoretically and computer simulations were performed.

The non-steady-state model~\cite{Kolegov201424} is based on one-dimensional
equations obtained based on mass and momentum conservation laws. The model
includes the convection--diffusion equation, the film thickness evolution
equation, and the equation of motion for a viscous fluid, all written in
cylindrical coordinates. The process is described in the initial stage when
the solution can be regarded as dilute. In~\cite{Kolegov201424}, the
capillary flow is taken into account, which is typical of a microcell
smaller than the capillary length. A subsequent modification of the
model~\cite{Kolegov2018344} also takes the hydrostatic pressure into account
(additional term in the equation of motion), which allows considering the
process in a macrocell. We note that the model in~\cite{Kolegov2018344} does
not use the lubrication approximation; the flow velocity is obtained from
the 1D equation of motion and $l\neq1$
in~\eqref{eq:ConservationLawForFilmThickness}, in~\eqref{eq:convection_diffusion},
and in the expression for the capillary pressure obtained by considering the
Young--Laplace equation and the curvature of an axisymmetric surface,
$$
P=-\sigma \left(\frac{1}{l^3}\frac{\partial^2h}{\partial r^2}+
\frac{1}{r l}\frac{\partial h}{\partial r}\right).
$$
This model thus accounts for the curvature of the two-phase boundary, which
is not always flat. The shape of the free surface of the liquid in the
microcell differs from that of the macrocell because of competition between
the capillary and gravitational forces. The approximation
$J=J_0[1-\cos(2\pi rN/R)]/2$ is used for the vapor flux density, which
agrees qualitatively with the numerical calculations~\cite{Harris2007}. Here,
$N$ is the number of circular holes in the mask, and $R$ is the cell radius. The
results of the calculations show that the flow carries particles into the
areas under the mask holes. Over time, the mass fraction of particles in
these areas increases. As the evaporation rate increases, the difference in
solution concentration in closed and open areas also increases. With slow
evaporation, the difference is smoothed. The reason for this is the
competition between diffusion and convection transfer.

The experiment in~\cite{Harris2007} with evaporation of a colloidal liquid
under a mask was simulated in~\cite{Tarasevich201626,Vodolazskaya2017}. The
two-dimensional model is based on the mass conservation law for a suspended
substance and a liquid.  The motion of the liquid--air boundary is set by
explicitly assuming that it is always flat. The equilibrium free boundary is
calculated based on an estimate of the evaporation rate. The height-averaged
radial flow velocity $\bar{u}(x,y)$ is determined from the mass conservation
law. A compensatory flow caused by evaporation is considered. The
convection--diffusion equation contains an additional term to account for
particle deposition (particle adhesion to the substrate). The vapor flux
density in~\cite{Tarasevich201626}
is described by the model law $J(r,\varphi)=J_0\sin^2(\pi r/R)
\sin^2(m\varphi/2)$, where the constant $m$ corresponds to the number of
symmetrically located holes in the mask above the drop.
The local vapor flux density $J$ at the two-phase boundary
in~\cite{Vodolazskaya2017} is determined using the solution of the Laplace
equation for the vapor concentration in the air area. The value of $J$ was
calculated for the initial film height. The dependence of this solution on
the further dynamics of the free surface was not considered. The effect on
the mass transfer of parameters such as the distance between the mask and
liquid, the ratio of the mask hole radius to the hole distance, the particle
deposition rate, the diffusion coefficient value, and the initial particle
fraction was considered in~\cite{Vodolazskaya2017}.

\begin{sidewaystable*}[htbp]
\tiny
\centering
\caption{Comparative characteristic of models}
\label{tab:ModelsComparison}
\begin{tabular*}{.9\textwidth}{|p{0.05\textheight}|p{0.07\textheight}|p{0.05\textheight}|p{0.08\textheight}|p{0.12\textheight}|p{0.06\textheight}|p{0.10\textheight}|p{0.05\textheight}|p{0.11\textheight}|p{0.07\textheight}|c|}
\cline{1-11}
\cline{1-11}
\parbox{7em}{Models vs.~characteristics} & Viscosity & Diffusion & Flow type & Vapor flux density & \parbox{7em}{Solidification in a liquid} & Heat transfer & \parbox{5em}{Other effects} & \parbox{12em}{Approaches and methods} & \parbox{8em}{Comparison with experiments} & Dimension \\
\cline{1-11}
\cline{1-11}
Routh \& Russel, 1998~\cite{RouthRussel1998} & constant & -- & Capillary flow, Darcy's law is partially taken into account. & Constant value. No flux in the masked area. & The liquid-solid interface is tracked explicitly in a film. & -- & -- & Lubrication approximation. Fourth-order Runge--Kutta method. & Qualitative comparison with~\cite{RouthRussel1998}. & 1D \\
\cline{1-11}
Deegan et al., 2000~\cite{Deegan2000} & Not taken into account explicitly, because there is no equation of motion. & -- & Compensatory flow from the mass conservation law. & Analytic approximation based on the solution of the Laplace equation for vapor taking the electrostatic analogy into account. & Single phase model. & -- & -- & Kinematic approach with an equilibrium droplet shape in which the flow velocity is determined from the conservation mass law. Analytic solution. & Qualitative comparison with~\cite{Deegan2000}. & 1D \\
\cline{1-11}
Fischer, 2002~\cite{Fischer2002} & constant & -- & Capillary flow. & Three model evaporation laws simulating the experiment~\cite{Deegan2000}. & Single phase model. & -- & -- & Lubrication approximation. Implicit Gear method. & Qualitative comparison with~\cite{Deegan2000}. & 1D \\
\cline{1-11}
Dietzel \& Poulikakos, 2005~\cite{Dietzel2005} & constant & Depends on particle concentration and temperature. & Thermocapillary flow. & Energy and mass balance analysis for the heated spherical droplets, taking the diffusion of vapor in air and the Stefan flow into account. & Single phase model. & Thermal diffusion in liquid and substrate. Laser heating of a free surface part. & Coagulation of particles. & Navier--Stokes equations written in Lagrange coordinates. FEM. & -- & 2D \\
\cline{1-11}
Yamamura et al., 2009~\cite{Yamamura2009} & Depends on particle concentration and temperature. & Depends on particle concentration. & Capillary flow. Marangoni convection. Surface tension depends on temperature and particle concentration. & A semi-empirical formula based on the difference in vapor pressure near the free surface of the film and away from it. & Single phase model. & Change in liquid and substrate temperature. Evaporative cooling. Heat transfer from the air. & Simulation of a moving substrate under periodically arranged nozzles. & Lubrication approximation. Implicit difference scheme on an nonuniform mesh. The discretized scheme was solved by the Thomas algorithm. & -- & 1D \\
\cline{1-11}
Vieyra Salas et al., 2012~\cite{VieyraSalas2012} & Depends on the solute concentration. & Depends on the solute concentration. & Flow caused by capillary and hydrostatic pressure gradients. Marangoni convection caused by the concentration and temperature gradients on the free surface. & Convection--diffusion equation for the vapor is solved. & The two-phase boundary is tracked implicitly. & Laser film heating. Cooling due to evaporation, radiation, and heat transfer to the air. Heat transfer in air and liquid heatsink. Substrate cooling with the heatsink. Thermal diffusion in a liquid and substrate. & -- & The steady-state heat transfer equation for the liquid in the heatsink and thermal conductivities of the substrate and the film. Hydro- and aerodynamic equations. ALE method. & Qualitative comparison with~\cite{VieyraSalas2012}. & 2D \\
\cline{1-11}
Arshad \& Bonnecaze, 2013~\cite{Arshad2013} & Depends on particle concentration. & Depends on particle concentration. & Flow caused by osmotic pressure gradient. & Depends on the thickness of the membrane. & The two-phase interface is tracked implicitly. & -- & -- & Navier--Stokes equations, convection--diffusion equation. FEM. & -- & 2D \\
\cline{1-11}
Vodolazskaya \& Tarasevich, 2017~\cite{Vodolazskaya2017} & -- & constant & Compensatory flow from the mass conservation law. & Solving the vapor diffusion equation in ambient air for the initial film height. & Single-phase model. & -- & Adhesion of particles to the substrate. & Equations based on the conservation mass law. Kinematic approach in flow velocity calculation. FEM. & Qualitative comparison with~\cite{Harris2007}. & 2D \\
\cline{1-11}
Hu et al., 2017~\cite{Hu2017} & -- & -- & Compensatory flow from the mass conservation law. & Linear approximation. & Precipitation over the contact line. & -- & Contact line movement. & Kinematic approach. Contact line dynamics is determined by the Onsager variational principle. & Qualitative comparison with~\cite{Deegan2000,ChenL2009,Pradhan2015}. & 2D \\
\cline{1-11}
Kolegov \& Lobanov, 2018~\cite{Kolegov2018344} & constant & const & Flow caused by the capillary and hydrostatic pressure gradient. & Depends on the spatial coordinate. Mimics the mask influence. & Single-phase model. & -- & -- & The equation of motion, the solution mass conservation law, the convection--diffusion equation. FDM. & -- & 1D \\
\cline{1-11}
Wedershoven et al., 2018~\cite{Wedershoven2018} & Depends on particle concentration. & constant & Capillary flow. & Depends on the vapor and particle concentrations near the free surface. & The two-phase boundary is tracked implicitly. & -- & -- & Hydro- and aerodynamic equations. ALE method. & Quantitative comparison with~\cite{Wedershoven2018}. & 2D \\
\cline{1-11}
Wedershoven et al., 2018~\cite{Wedershoven201892} & Depends on particle concentration. & constant & Capillary \& solutal Marangoni flow & Depends on particle concentration. & The two-phase boundary is tracked implicitly. & -- & Substrate motion. & Lubrication approximation. FEM. & Qualitative comparison with~\cite{Wedershoven201892}. & 1D \\
\cline{1-11}
Kolegov, 2018~\cite{Kolegov2018113} & Depends on the particle concentration. & constant & Capillary flow. & Depends on temperature, particle concentration, and coordinate (mimics mask effect). & The two-phase boundary is tracked implicitly. & External lamp heating. Heat transfer in a liquid. Thermal conductivity in a substrate. Heat exchange between the liquid and substrate. Evaporative cooling. & -- & Equations based on the mass and energy conservation laws. Flow velocity calculation via lubrication approximation. FDM. & Qualitative comparison with~\cite{Routh2011}. & 1D \\
\cline{1-11}
\cline{1-11}
\end{tabular*}
\end{sidewaystable*}

The appearance of asymmetric precipitation during the evaporation of two
adjacent droplets was studied in~\cite{Hu2017}. A similar experiment
was performed previously, for example, in ~\cite{Deegan2000}. The simple
model of a thin drop proposed in~\cite{Hu2017} describes a process in a
two-dimensional region parallel to the substrate. The shape of the drop is
determined by a parabolic approximation. The height-averaged radial velocity
is found from the solution mass conservation
law~\eqref{eq:ConservationLawForFilmThickness}. The time dependence
of the droplet base radius is determined based on the Onsager variational
principle. A linear approximation is used for the nonuniform vapor flux
density. The initial particle positions are randomly generated from a
uniform distribution. Only the lateral flow transfer is taken into account.
Depending on the parameters controlling the evaporation and friction of the
contact line, the model~\cite{Hu2017} predicts fan-like and eclipse-like
deposition patterns.

A mathematical model was developed in~\cite{Wedershoven2018} for the active
method of evaporative lithography where a source of dry airflow in the form
of a concentric nozzle is located above the film (see Sec.~\ref{sec:experiments}).
The problem was solved in two domains, namely, in a liquid film and in the
air above it. Steady-state equations were used for the air domain. The
model in~\cite{Wedershoven2018} describes the airflow velocity using the
equation of motion. The vapor concentration is described by the
convection--diffusion equation. Solving these equations allows calculating
the vapor flux density on the free surface of the film. In addition, when
calculating $J$, the authors~\cite{Wedershoven2018} roughly considered its
dependence on the particle concentration at the two-phase boundary. The
nonsteady equations were used to describe mass transfer in the film. In the
equation of motion, viscosity was taken out of the derivative, but in this
case, its dependence on the particle concentration was described (and
depends implicitly on space coordinate). The particle transfer was described
by the convection--diffusion equation, where the diffusion coefficient is
assumed to be constant. The ALE method was used in~\cite{Wedershoven2018} to
solve the problem. The dynamics of film thickness and particle concentration
in space and time were described using numerical simulations. The final
thickness of the solid film for different values of dry air flow rate from
the concentric nozzle was obtained. The numerical results agreed
qualitatively (in some cases quantitatively) with the experimental
results~\cite{Wedershoven2018}.

In~\cite{Wedershoven201892}, a 1D model based on the lubrication
approximation was used to justify the experimental method mathematically
(see Sec.~\ref{subsec:HybridMethods}). The expression for the
height-averaged flow velocity takes the effect of capillary pressure,
solutal Marangoni flow, and substrate motion into account. The viscosity
depends on the concentration of the solution and is described empirically.
The evaporation rate depends linearly on the concentration of the polymer
solution in the model~\cite{Wedershoven201892}.

As a part of the analysis of the state-of-the-art in this direction,
Table~\ref{tab:ModelsComparison} contains comparative characteristics of the
models. Further developments in modeling of the formation of micro- and
nanostructures with nonuniform evaporation of droplets and films will allow
a better understanding the basic mechanisms, finding out how to manage the
process, and obtaining particle deposit forms with greater spatial accuracy.

\section{Conclusions and outlooks}
\label{sec:conclusion}

Evaporative lithography is a new, actively developing field. This type of
lithography is based on generating ambient conditions for nonuniform
evaporation along the free surface of a droplet or film. The resulting
convection flows can be caused by different forces, but their initial cause
in any case is evaporation. For example, compensatory flows caused by the
deviation of the free surface shape from the equilibrium position can be
induced by both capillary and gravitational forces. Nonuniform evaporation
often results in a variation of fluid temperature and solution concentration
on the free surface. This results in a surface tension gradient, causing
Marangoni flows to emerge. If temperature and concentration gradients occur
in the bulk, then Rayleigh-B\'enard flows arise associated with
the heterogeneity of density. Other mechanisms of flow generation are also
possible that eventually carry the dissolved substance or particles
suspended in liquids to the required local areas, forming spatially
patterned deposits in a controlled manner.

Evaporative lithography methods can be divided into active and passive. In
passive methods, the key parameter values are set initially (the mask hole
size, the liquid--mask gap width, the initial particle concentration or film
thickness, the distance between the holes, the geometry of the membrane,
etc.). In contrast to passive methods, active methods are characterized by
the presence of key parameters that can be adjusted in real time (radiation
power, airflow velocity, substrate motion speed, etc.). This allows
dynamically influencing the geometry of the pattern being formed. Hybrid
methods that combine evaporative lithography with other EISA or non-EISA
approaches can be identified as a separate subgroup. This expands the
possibilities of evaporative lithography, which stimulates active research
in this area. The great interest in evaporative lithography is driven by
many promising applications in various fields. Let us mention just a few of
them. The creation of transparent flexible conductive films and organic
transistors is extremely important for optical and microelectronics. In
nanotechnologies, special attention is paid to functional coatings where the
obtained thin-film pattern has the required physical and chemical
properties. New miniature instruments are created for health care
development allowing express-diagnostics and delivery of drugs to particular
parts of the human body.

Along with experimental studies, mathematical models are being developed and
improved. Theoretical and numerical studies facilitate understanding of the
physical processes occurring in evaporative lithography. Modeling allows
identifying key system parameters and facilitates the determination of the
range of parameter values for efficient control over patterns formation with
the required geometric and physicochemical characteristics. Currently, there
are several models describing the process of heat and mass transfer in
evaporative lithography. In most cases, these are one-dimensional and/or
single-phase models that take only individual processes at a qualitative
level into account. Development in this direction is relatively slow because
the phenomena and processes observed in evaporative lithography are complex.

As a rule, the height of formed sediments and hard coverings varies from one
micron~\cite{HarrisLewis2008,Parneix2010,Anyfantakis2017} to several
tens~\cite{RouthRussel1998,Harris2007,Routh2011} or hundreds of
microns~\cite{Mansoor2011}. In a plane parallel to the substrate surface,
the size of individual structure elements ranges from hundreds of
microns~\cite{Layani2009,Mansoor2011,Yen2018} to several
millimeters~\cite{RouthRussel1998,Utgenannt2013,Cavadini2015,Zhao2016}.
In this respect, evaporative lithography is inferior to some other EISA
methods, which allow obtaining stripes of 10~microns or less in
width~\cite{Li2014,Li2018124}. Further development of evaporative
lithography will allow obtaining patterned deposits and solid patterned
films with more accurate spatial resolution. This will allow forming
non-blurred patterns on the substrate. Such micro- and nanostructures may be
further needed for new applications in the biosensors, labs-on-chip,
functional coatings, and many others. For this, the existing passive and
active methods must be improved. New methods must be also developed. A
hybrid approach is most likely to achieve significant results.

It is hoped that further experimental studies of evaporative lithography will
help resolve some of the existing technical difficulties. For example, one
problem that deserves attention is the need to increase an area of a formed
coating. This area is usually now measured in square millimeters or
centimeters, but such dimensions are insufficient for some applications. An
example is the problem of microstructuring the surface a boat hull to obtain
a coating that prevents the growth of marine organisms~\cite{Efimenko2009}.
Further development of theoretical studies and mathematical modeling of
evaporative lithography seems especially important in terms of improving the
control over the pattern forming process. We believe that the development of
more complete and hence more complex mathematical models requires a special
focus. This primarily relates to multiphase and multidimensional
simulations. Matters of interest include liquid rheology, flows in porous
media, drop or film solidification (its phase transition into glass, gel,
melt, etc.), thermal, physical, and chemical properties of the substrate
(e.g., composite surfaces or porous wafers), and binary particle systems
(including those with versatile shapes and properties).

\section{Acknowledgment}
This work is supported by the grant 18-71-10061 from the Russian
Science Foundation.

\bibliographystyle{elsarticle-num}
\bibliography{refs}
\end{document}